\newcommand{\chirpmass}{\ensuremath{\mathcal{M}}}
\newcommand{\healpy}{\tt\string healpy}

\newcommand{\mosfit}{\tt\string MOSFiT}
\newcommand{\tm}{\tt{Treasure Map}}

\newcommand{\qcr}{\fontfamily{qcr}\selectfont}

\documentclass[twocolumn]{aastex631}

\usepackage{hyperref}
\usepackage{natbib}

\begin{document}

\title{The Potential of Coordinated Gravitational-Wave Followup for Improving Kilonova Detection Prospects: Lessons from GW190425}

\author[0000-0002-0675-0887]{Ido Keinan}
\affiliation{The School of Physics and Astronomy, Tel Aviv University, Tel Aviv 69978, Israel}

\author[0000-0001-7090-4898]{Iair Arcavi}
\affiliation{The School of Physics and Astronomy, Tel Aviv University, Tel Aviv 69978, Israel}

\correspondingauthor{Ido Keinan}
\email{idokeinan1@mail.tau.ac.il}

\begin{abstract}
The discovery of a kilonova associated with the GW170817 binary neutron star merger had far-reaching implications for our understanding of several open questions in physics and astrophysics. Unfortunately, since then, only one robust binary neutron star merger was detected through gravitational waves, GW190425, and no electromagnetic counterpart was identified for it following an uncoordinated search. In order to estimate the potential difference that coordinated followup could make for future events, we analyze all reported electromagnetic followup observations of GW190425. We find that even for a large gravitational-wave localization uncertainty, such as this one, most of the 90\% probability region can be covered within hours with a coordinated search, given the observational resources expended in this case by the community. However, more than 5 days after the GW190425 merger, its uncoordinated search covered only 50\% of the probability, with some areas observed over 100 times, and some never observed. According to some models, the GW190425 kilonova could have been detected, despite the larger distance and higher component masses compared to GW170817. These results emphasize that coordinated followup of gravitational-wave events can, in principle, significantly improve both the chances of finding electromagnetic counterparts, and the time it takes to do so, compared to uncoordinated searches.
\end{abstract}

\section{Introduction}
\label{sec:intro}

Binary neutron star (BNS) mergers were predicted to produce both gravitational waves \citep[GWs;][]{Clark1977} and electromagnetic (EM) radiation. The latter was theorized to be emitted in the form of two types of transients: a short gamma-ray burst \citep[GRB;][]{Eichler1989}, and longer-duration emission from the radioactive decay of heavy elements synthesized via the $r$-process in the ejecta \citep{li_paczynski_kilonova}. This transient has been nicknamed ``macronova'' \citep{Kulkarni2005} or ``kilonova'' \citep{metzger_kilonova}\footnote{Hereafter we use the term ``kilonova'' rather than ``macronova'' because it has been more widely adopted by the community.}.

All three predictions were confirmed with the first discovery of a GW signal from a BNS merger, GW170817, detected and localized by the Advanced Laser Interferometer Gravitational-wave Observatory \citep[LIGO;][]{ligo_detectors}, and the Advanced Virgo detector \citep{virgo_detector}, during their second observation run (O2), on August 17, 2017 \citep{gw170817}. The GW source was localized to a region of $\sim30$ deg$^{2}$ at a distance of $\sim40$ Mpc and had a total binary mass of $2.74^{+0.04}_{-0.01}$ M$_{\odot}$ \citep{gw170817}.
A short GRB, GRB170817A, consistent with the GW localization, was detected $\sim$2 seconds after the GW-determined merger time \citep{fermi_integral_grb170817a} by the Fermi Gamma-ray Burst Monitor \citep[Fermi-GBM;][]{fermi_gbm, goldstein2017_grb170817a}, and the International Gamma-Ray Astrophysics Laboratory \citep[INTEGRAL;][]{integral, savchenko2017_grb170817a} SPectrometer on INTEGRAL Anti-Coincident Shield (SPI-ACS).
An optical transient, AT 2017gfo, also consistent with the GW localization, was detected 11 hours later \citep[e.g.][]{coulter_gw170817_swope, gw170817_multi_messenger}.

Specifically, AT 2017gfo was consistent with kilonova predictions whereby at the coalescence of a BNS system, $10^{-4}-10^{-2}$ M$_{\odot}$ of neutron-rich material are ejected at velocities of $0.1-0.3c$ in the equatorial plane due to tidal effects \citep[e.g.][]{rosswog_mass_ejection, hotokezaka_mass_ejection}. Additional mass was predicted to be ejected in the polar direction from the contact region at the time of the merger \citep[e.g.][]{bauswein_mass_ejection, hotokezaka_mass_ejection}. Lanthanides formed in low electron-fraction material, such as the equatorial tidal ejecta, have a high opacity \citep{kasen_opacity, tanaka_hotokezaka_opacity}, making the light curve redder, fainter, and longer lived compared to low lanthanide ejecta \citep{barnes_kasen_opacity, grossman_opacity}, such as that from the polar regions. The result, seen in AT 2017gfo, is a kilonova of at least two components: one blue and short-lived from the lanthanide-poor ejecta, and one red and longer-lived from the lanthanide-rich ejecta. For a review of this event, see for e.g. \cite{Nakar2020} and \cite{margutti2021}. 

GW170817 and its EM counterparts confirmed that BNS mergers are sites of $r$-process nucleosynthesis \cite[e.g.][]{kasen_models}, and confirmed the connection between short GRBs and BNS mergers. This event also demonstrated a novel method to constrain the Hubble constant \citep{abbott_hubble_const}. However, many open questions regarding the properties of the kilonova emission remain, such as whether the source of the early blue emission is indeed a distinct lanthanide-poor ejecta component \citep[other physical mechanisms were proposed, e.g.][]{kasliwal_gw170817_obs,Piro2018,waxman2018}. More BNS detections and multi-wavelength observations of their EM counterparts, especially during the first few hours \citep[e.g.][]{Arcavi_2018} are needed to answer these questions.

\begin{figure} 
\includegraphics[width=0.5\textwidth]{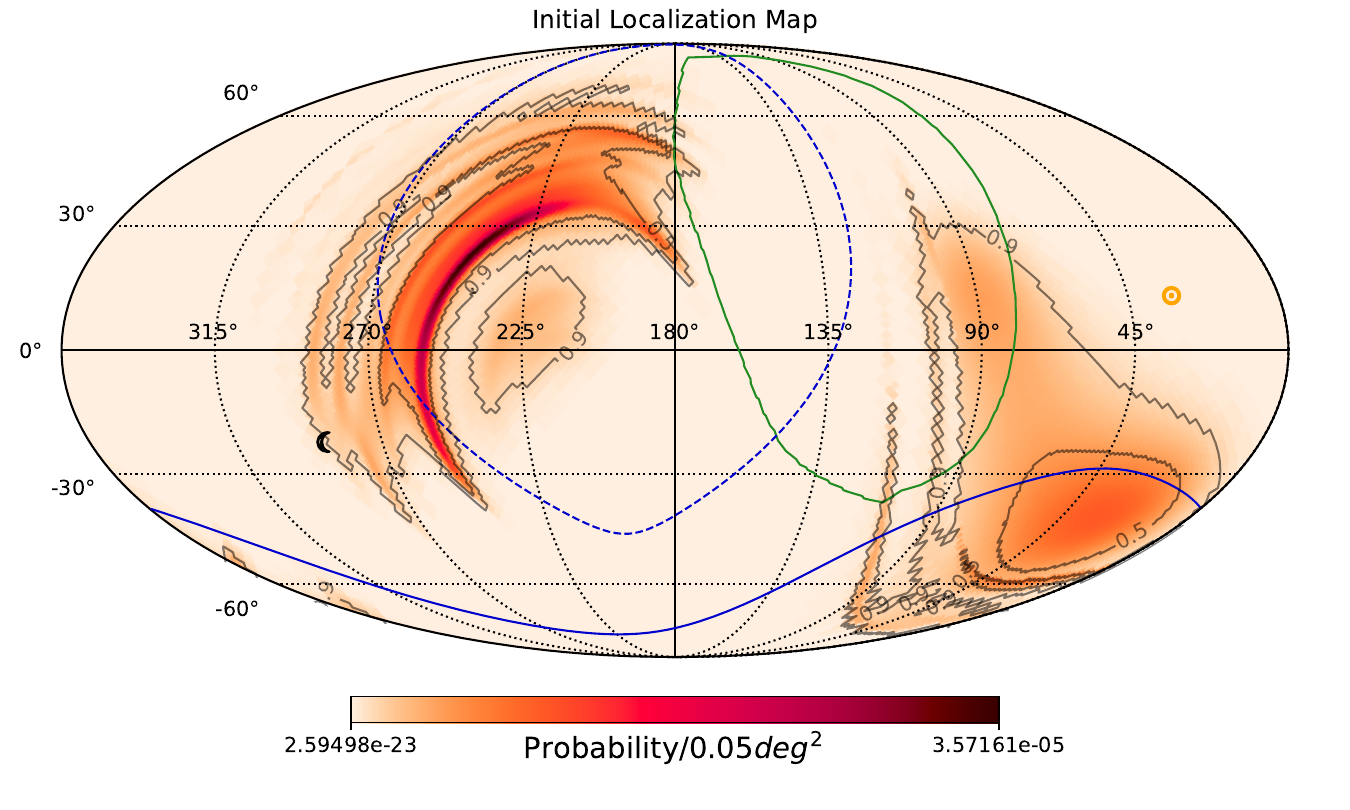}
\includegraphics[width=0.5\textwidth]{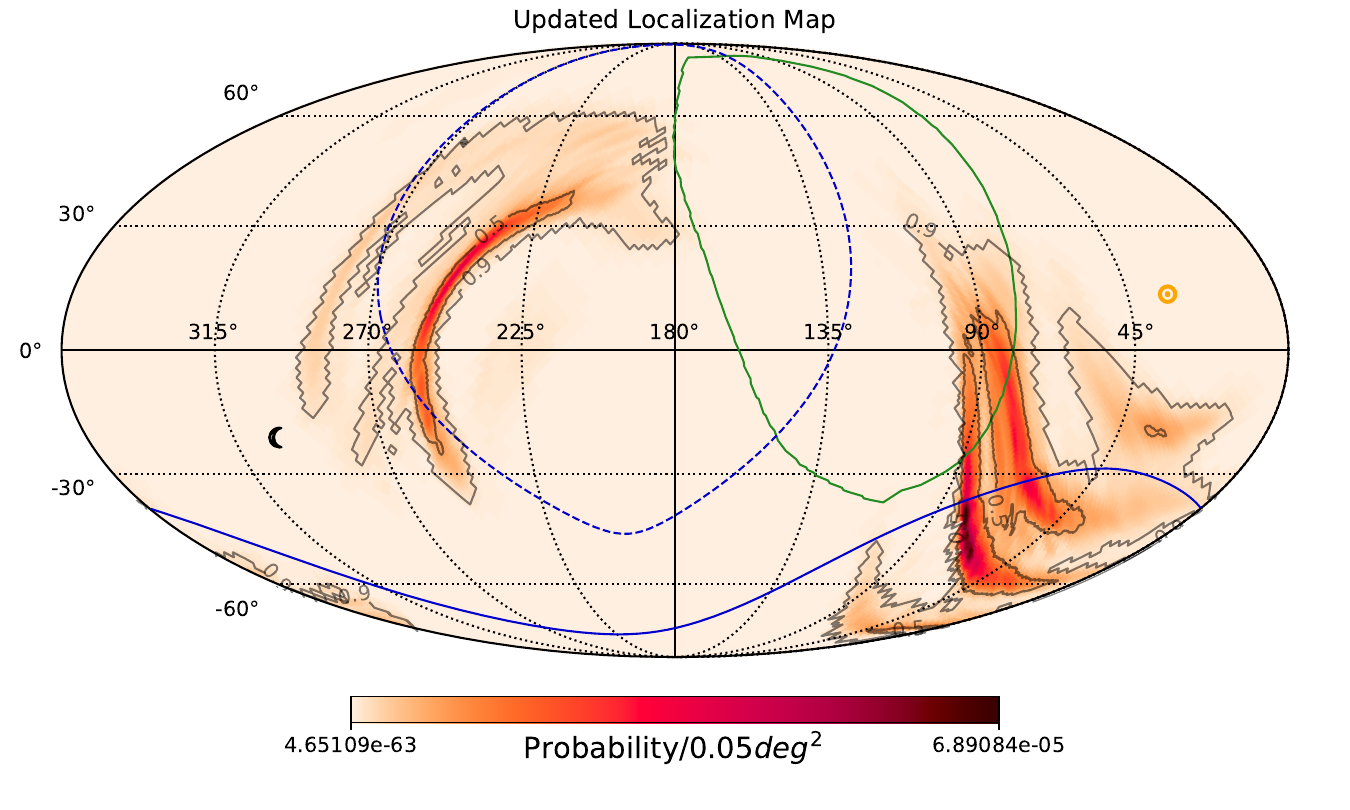}
\includegraphics[width=0.5\textwidth]{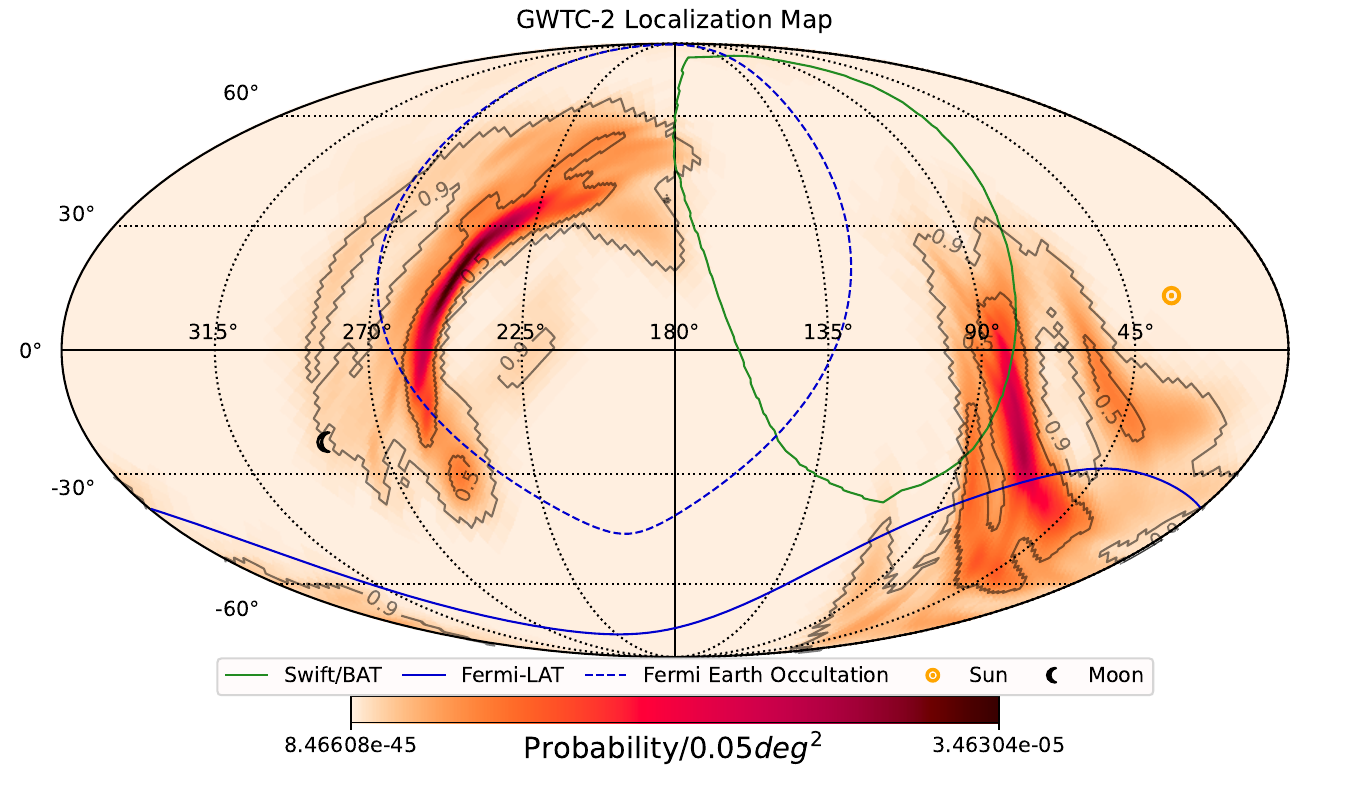}
\caption{GW190425 localizations released in the initial (top) and update (middle) alerts, and in the final GWTC-2 publication \citep[bottom;][]{gwtc2}. White and grey contours denote the regions containing 50\% and 90\% of the probability of the source location, respectively. The Swift/BAT and Fermi-LAT footprints at the time of the merger are denoted by the green solid line and below the blue solid line, covering areas of 4596 deg$^{2}$ and 8602 deg$^{2}$, respectively. The Earth occultation of 14,187 deg$^{2}$ for the all sky Fermi-GBM instrument is shown in a blue dashed line. The position of the Sun and Moon at the time of the merger are denoted in their respective symbols.}
\label{fig:localizations}
\end{figure}

The second (and so far, last) robust GW detection of a BNS merger occurred at the start of the third GW-detector observing run (O3). GW190425 was discovered on April 25, 2019, at 08:18:05 UTC, at a distance of $159^{+69}_{-72}$ Mpc and with a total binary mass of $3.4^{+0.3}_{-0.1}$ M$_{\odot}$ \citep{gw190425}. The event was detected only by a single LIGO detector, making its localization poorly constrained, with a 10183  deg$^{2}$ 90\% uncertainty region initially \citep{GCN24168_ligo}. This region shrank about one day later to 7461 deg$^{2}$ \citep{GCN24228_ligo}, but grew back to 9881 deg$^{2}$ in the final localization published in the second GW Transient Catalog \citep[GWTC-2;][]{gwtc2}. 
The localization parameters are summarized in Table \ref{tab:alerts_summary}, and the localization maps are shown in Figure \ref{fig:localizations}. 

\begin{deluxetable}{lllll} 
\label{tab:alerts_summary}
\tablecaption{GW190425 localization areas and distances.}
\tablehead{\colhead{Epoch} & \colhead{Date} & \colhead{90\% Area} & \colhead{50\% Area} & \colhead{Distance}
\\[-0.2cm]
\colhead{} & \colhead{[UTC]} & \colhead{[deg$^{2}$]} & \colhead{[deg$^{2}$]} & \colhead{[Mpc]}}
\startdata
Initial & 2019-04-25 & 10,183 & 2806 & $154\pm45$ \\ Alert & 08:19:27 & & & \\ \hline 
Update & 2019-04-26 & 7461 & 1378 & $156\pm41$ \\ Alert & 10:48:17 & & & \\ \hline 
GWTC-2 & 2020-07-27 & 9881 & 2400 & $157\pm43$ \\ & 16:41:49
\enddata
\tablecomments{``90\% (50\%) Area'' refers to the sky area that contains 90\% (50\%) of the probability for the location of the source.}
\end{deluxetable}

An extensive EM followup effort to find the kilonova associated with GW190425 began shortly after the initial alert. However, no significant kilonova candidate was identified. This was attributed to the larger distance, higher component masses, more equal mass ratio, and to the larger localization compared to GW170817 \citep[e.g.][]{hosseinzadeh_190425_190426,lundquist_saguaro_non_detection_190425,gw190425,  foley_updated_gw190425_parameters, gompertz_goto_o3,  kyutoku_190425_as_bhns, sagues_o3_non_detection, camilletti_nr_sim, dudi_nr_simulations, radice_no_prompt_collapse}. However, here we show that, according to some models, the emission from the GW190425 kilonova could have been detected by facilities that took part in the search, and that in principle most of the localization could have been covered within hours, had the EM followup effort been coordinated.

Coordinating EM followup of GW events is challenging. Reports of observations and possible kilonova candidates during O2 and O3 were primarily communicated through GRB Coordinates Network\footnote{\url{https://gcn.nasa.gov/}} (GCN; now the General Coordinates Network) circulars.
GCN circulars were designed to be human readable, with no unified text format. As such they are not ideal when reporting lists of coordinates, bands, and sensitivities observed. The non machine-readable format makes it very difficult to parse, store and interpret such information in real time. 

One tool designed to address the challenge of coordinating EM followup of GW events is the  {\tm}\footnote{\url{https://treasuremap.space/}} \citep{treasure_map}. {\tm} is a web application that collects, stores and distributes followup reports through a machine-friendly Application Programming Interface (API). The API allows observers to report their planned and executed observations and to retrieve those of other facilities in order to inform their own counterpart search plan in real time. In addition, an interactive visualization allows observers to see their observations and those of other facilities on a map, together with GW localization in an intuitive way. All of the reports are also available to download at a later date for analyzing followup efforts and non-detection statistics. Unfortunately, the {\tm} did not exist during GW190425, leading to very inefficient followup, as we will show here. Our goal is to encourage the community to use tools such as the {\tm} during the next BNS merger, in order to increase the chances of finding the EM counterpart, and to do so as quickly as possible. We do so by demonstrating the impact of uncoordinated observations on missing the counterpart to GW190425.

In this work, we analyze all reported infrared, optical and ultraviolet followup efforts of GW190425. In Section \ref{sec:data} we describe our data sources and collection process; in Section \ref{sec:analysis}, we analyze the sky coverage and depth achieved in the context of the expected kilonova emission from this event; and, finally, in Section \ref{sec:conclusions}, we discuss our conclusions.

\section{Data}
\label{sec:data}
In order to analyze the efficiency of the search for the GW190425 EM counterpart we collected pointing and photometric data of infrared, optical and ultraviolet observations obtained as followup of GW190425 up to 2019 April 30, 23:57:21 UTC (5.65 days after the merger)\footnote{Reports are available up to approximately $\sim$15 days after the merger, though they become increasingly sparse over time. We selected this cut-off as it encompasses about 95\% of the reports, while avoiding the long tail of sparse reporting at later times.}, which were reported in one of the following sources (removing duplicate reports of the same observations): The {\tm} (1419 observations reported retroactively), 122 GCN circulars (1339 observations), the MASTER global robotic net \citep[MASTER-Net;][]{master_robotic_net} website\footnote{\url{http://observ.pereplet.ru/}} (4491 observations), the \cite{coulter_analysis_2024} report on the One-meter Two-hemispheres (1M2H) followup (345 observations), the \cite{smartt_ps_atlas} report of the Panoramic Survey Telescope and Rapid Response System (Pan-STARRS) and the Asteroid Terrestrial-impact Last Alert System (ATLAS) followup (232 and 422 observations, respectively), and the \cite{ddoti_observations} report (augmented with R.L. Becerra, private communication) of the Deca-Degree Optical Transient Imager (DDOTI) followup (10 observations).

In total, 8258 unique observations were collected, including time stamps and bands used. Limiting magnitudes are available for all except for the pointings of the Corrector de Optica \'{A}ctiva y de Tilts al L\'{ı}mite de dIfracci\'{o}n \citep[COATLI;][]{2019GCN.24239....1W}, the Korea Microlensing Telescope Network \citep[KMTNet;][]{2019GCN.24216....1K} and Pan-STARRS in the the $z$-band, which we include here for coverage analysis only. Instrument footprints and fields of view (FOVs) for all pointings analyzed here were taken from the {\tm} when available, from GCN reports when provided there, and from instrument-specific websites and publications otherwise (all references are provided in Table \ref{tab:observations_summary}). Some galaxy-targeted searches provided target galaxy names rather than exact coordinates pointed to. In such cases, we obtained the galaxy coordinates from the Galaxy List for the Advanced Detector Era (GLADE) catalog \citep{dalya_glade2p3} or the SIMBAD Astronomical Database \citep{simbad} and assumed the target galaxies were positioned at the center of the observed footprint. A summary of the observations and instruments used in this analysis is presented in Table \ref{tab:observations_summary}, and the full list of all pointings gathered is presented in Table \ref{tab:all_obs}.

\startlongtable
\begin{deluxetable*}{lccccc} 
\label{tab:observations_summary}
\tablecaption{GW190425 EM followup observations reported in circulars, the {\tm}, by MASTER-Net on their website, and in \cite{coulter_analysis_2024}, \cite{smartt_ps_atlas} and \cite{ddoti_observations}, out to 5.65 days after merger.}
\tablewidth{0pt}
\tablehead{\colhead{Group/Facility/Instrument} & \colhead{No. of} & \colhead{Bands} & \colhead{Field of View} & \colhead{Median $5\sigma$ Depth} & \colhead{Pointings Reference}\\
\colhead{} & \colhead{Pointings} & \colhead{} & \colhead{[deg$^{2}$]} & \colhead{[AB Mag]} & \colhead{}}
\startdata
MASTER-Net$^{a}$ & 4491 & Clear & 8.00 & 19.10 & 1 \\
ZTF$^{b}$ & 596 & g,r,i & 46.73 & 21.05 & 2, 3 \\
ATLAS$^{c}$ & 422 & o,c & 28.89 & 19.26 & 4 \\
Swift/UVOT$^{d}$ & 392 & u & 8.0e-02 & 19.40 & 2, 5 \\
GOTO$^{e}$ & 303 & V,g & 18.85 & 19.85 & 2, 6 \\
Pan-STARRS$^{f}$ & 232 & w,i,z & 7.06 & 21.58 & 4 \\
J-GEM/Subaru$^{g}$ & 154 & r & 2.0e-03 & 24.00 & 7 \\
1M2H/Swope$^{h}$ & 151 & B,V,u,g,r,i & 0.25 & 21.55 & 8 \\
COATLI$^{i}$ & 128 & w & 3.3e-02 & n/a & 9 \\
KMTNet 1.6m$^{j}$ & 120 & R & 4.00 & n/a & 10 \\
KAIT$^{k}$ & 101 & Clear & 1.2e-02 & 19.00 & 11 \\
Harold Johnson Telescope/RATIR$^{l}$ & 99 & J,H,Y,Z,g,r,i & 8.1e-03 & 20.74 & 12, 13 \\
GROWTH/LOT$^{m}$ & 93 & R,g,r,i & 3.6e-02 & 19.79 & 14, 15, 16 \\
1M2H/Nickel$^{n}$ & 93 & r,i & 1.1e-02 & 20.24 & 8 \\
MMTO/MMTCam$^{o}$ & 81 & g,i & 2.0e-03 & 21.48 & 2, 17, 18 \\
GWAC-F60A$^{p}$ & 80 & Clear & 9.0e-02 & 18.97 & 19 \\
CNEOST$^{q}$ & 75 & R & 9.00 & 19.86 & 20 \\
BOOTES-5$^{r}$ & 63 & Clear & 2.78 & 20.50 & 21 \\
SAGUARO/CSS 1.5m$^{s}$ & 61 & Clear & 5.00 & 21.30 & 2 22, 23 \\
LCO 1m$^{t}$ & 58 & g,r,i & 0.20 & 21.30 & 24, 25, 26, 27 \\
1M2H/Thacher$^{u}$ & 53 & g,r,z & 0.12 & 19.55 & 8 \\
GRANDMA/LesMakes T60$^{v}$ & 52 & Clear & 17.64 & 19.20 & 28 \\
GRANDMA/AZT$-$8$^{w}$ & 52 & B,R & 0.25 & 19.00 & 28 \\
GRANDMA/Abastumani T70$^{x}$ & 40 & R & 0.25 & 16.29 & 28 \\
Xinglong 60/90cm$^{y}$ & 35 & Clear & 2.25 & 18.00 & 29 \\
McDonald Observatory 2.1m/CQUEAN$^{z}$ & 30 & i & 6.1e-03 & 20.44 & 30 \\
1M2H/ANDICAM-CCD$^{aa}$ & 27 & I & 1.1e-02 & 20.70 & 8 \\
GROWTH/Kitt Peak 2.1m/KPED$^{ab}$ & 25 & I,g,r & 7.0e-03 & 20.40 & 31, 32, 33 \\
YAHPT$^{ac}$ & 23 & R & 3.4e-02 & 18.18 & 34 \\
1M2H/ANDICAM-IR$^{ad}$ & 21 & J,H,K & 1.5e-03 & 13.45 & 8 \\
Liverpool Telescope/IO:O$^{ae}$ & 19 & g,r,i,z & 2.8e-02 & 22.24 & 35, 36 \\
LOAO 1m$^{af}$ & 13 & R & 0.22 & 19.23 & 37 \\
GROWTH-India$^{ag}$ & 10 & g,r,i & 0.49 & 18.41 & 38, 39 \\
Konkoly 0.8m$^{ah}$ & 10 & g,r & 9.0e-02 & 20.40 & 40 \\
DDOTI$^{ai}$ & 10 & w & 11.56 & 19.95 & 41 \\
IRSF 1.4m/SIRIUS$^{aj}$ & 7 & J,H,K & 1.6e-02 & 15.31 & 42, 43 \\
TAROT TRE$^{ak}$ & 7 & R & 17.64 & 17.23 & 44 \\
Konkoly 0.6/0.9m$^{al}$ & 5 & Clear & 1.36 & 21.50 & 40 \\
Lijiang 2.4m$^{am}$ & 5 & g,r & 2.56 & 19.04 & 45 \\
GRANDMA/CAHA 2.2m$^{an}$ & 5 & U,B,V,R,I & 4.5e-02 & 22.30 & 46 \\
DECam$^{ao}$ & 4 & g,r,i,z & 3.35 & 23.85 & 47 \\
MPG/GROND$^{ap}$ & 3 & r,i,z & 8.1e-03 & 20.00 & 48 \\
NOT/ALFOSC$^{aq}$ & 3 & g,r,i & 1.1e-02 & 20.39 & 49 \\
OSN 1.5m$^{ar}$ & 2 & R & 6.8e-02 & 21.30 & 50, 51 \\
ANU/SkyMapper$^{as}$ & 2 & i & 5.62 & 20.00 & 52 \\
GTC/OSIRIS$^{at}$ & 1 & r & 1.8e-02 & 22.80 & 51 \\
VISTA/VIRCAM$^{au}$ & 1 & K & 2.14 & 19.55 & 53 
\enddata

\tablecomments{Sources which did not provide their observation depths are marked with ``n/a'' in the depth column. The median depth for Pan-STARRS is for the $w$ and $i$ bands, as no observation depths were provided for the $z$-band.}
\tablenotetext{}{Instrument references: 
$^a$Mobile Astronomical System of TElescope Robots \citep{master_robotic_net}; 
$^b$Zwicky Transient Facility \citep{zwicky_transient_facility};
$^c$Asteroid Terrestrial-impact Last Alert System \citep{atlas}
$^d$Swift Ultraviolet/Optical Telescope \citep{swift_uvot};
$^e$Gravitational-wave Optical Transient Observer \citep{goto};
$^f$Panoramic Survey Telescope and Rapid Response System \citep{pan_starrs}
$^g$Japanese Collaboration for Gravitational-wave ElectroMagnetic followup \citep{subaru_1, subaru_2};
$^h$One-Meter Two-Hemispheres/Swope \citep{swope};
$^i$Corrector de \'{O}ptica \'{A}ctiva y de Tilts al L\'{\i}mite de dIfracci\'{o}n \citep{coatli};
$^j$Korea Microlensing Telescope Network \citep{kmtnet};
$^k$Katzman Automatic Imaging Telescope \citep{kait};
$^l$Reionization and Transients Infrared Camera \citep{ratir};
$^m$Global Relay of Observatories Watching Transients Happen/Lulin One-meter Telescope \citep{growth};
$^n$Nickel Direct Imaging Camera (\url{https://mthamilton.ucolick.org/techdocs/instruments/nickel_direct/intro/});
$^o$MMT Observatory \citep{mmtcam};
$^p$Ground Wide Angle Cameras Array \citep{gwac_a};
$^q$Chinese Near Earth Object Survey Telescope;
$^r$Burst Observer and Optical Transient Exploring System 5;
$^s$Searches After Gravitational-waves Using ARizona's
Observatories/Catalina Sky Survey \citep{css};
$^t$Las Cumbres Observatory \citep{las_cumbres_observatory};
$^u$Thacher ACP Camera \cite{thacher};
$^v$Global Rapid Advanced Network Devoted to the Multi-messenger Addicts \citep{grandma};
$^w$Astronomical Reflecting Telescope 8 \citep{grandma};
$^x$Abastumani\citep{grandma};
$^y$Xinglong 60/90cm, Chinese Academy of Sciences;
$^z$Camera for QUasars in EArly uNiverse \citep{cquean};
$^{aa}$A Novel Dual Imaging CAMera (ANDICAM) CCD \citep{andicam};
$^{ab}$Kitt Peak EMCCD Demonstrator;
$^{ac}$YAoan High Precision Telescope \citep{yahpt};
$^{ad}$ANDICAM-IR \citep{andicam};
$^{ae}$Infrared-Optical:Optical \citep{liverpool_telescope};
$^{af}$Lemmonsan Optical Astronomy Observatory \citep{loao1m};
$^{ag}$\cite{growth};
$^{ah}$Konkoly 0.8m Telescope, Research Centre for Astronomy and Earth Sciences;
$^{ai}$Deca-Degree Optical Transient Imager \citep{ddoti};
$^{aj}$InfraRed Survey Facility/Simultaneous InfraRed Imager for Unbiased Survey \citep{irsf_sirius};
$^{ak}$Télescope à Action Rapide pour les Objets Transitoires \citep{tarot};
$^{al}$Konkoly 0.6/0.9m (\url{https://old.konkoly.hu/konkoly/telescopes.html});
$^{am}$\citep{lijiang};
$^{an}$Centro Astronómico Hispano en Andalucía \citep{grandma};
$^{ao}$Dark Energy Camera \citep{dark_energy_camera};
$^{ap}$Max Planck Gesellschaft/Gamma-Ray burst Optical and Near-infrared Detector \citep{grond};
$^{aq}$Nordic Optical Telescope/ALhambra Faint Object Spectrograph and Camera \citep{not};
$^{ar}$Observatorio de Sierra Nevada (\url{https://www.osn.iaa.csic.es/en/page/15-m-telescope});
$^{as}$Australian National University \citep{skymapper};
$^{at}$Gran Telescopio Canarias/Optical System for Imaging and low-intermediate Resolution Integrated Spectroscopy \citep{osiris};
$^{au}$Visible and Infrared Survey Telescope for Astronomy/VISTA InfraRed CAMera \citep{vista, vircam}.}
\tablenotetext{}{Pointings references:
$^1$\url{https://observ.pereplet.ru/} \citep{master_robotic_net};
$^2$\url{https://treasuremap.space/} \citep{treasure_map};
$^3$\cite{coughlin2019_ztf};
$^4$\cite{smartt_ps_atlas};
$^5$\cite{oates2021_swift_uvot_190425};
$^6$\cite{gompertz2020_goto_190425};
$^7$\cite{2019GCN.24192....1S};
$^8$\cite{coulter_analysis_2024};
$^9$\cite{2019GCN.24239....1W};
$^{10}$\cite{2019GCN.24216....1K};
$^{11}$\cite{2019GCN.24179....1Z};
$^{12}$\cite{2019GCN.24238....1B};
$^{13}$\cite{2019GCN.24335....1T};
$^{14}$\cite{2019GCN.24193....1T};
$^{15}$\cite{2019GCN.24274....1T};
$^{16}$\cite{2019GCN.24303....1K};
$^{17}$\cite{hosseinzadeh_190425_190426};
$^{18}$\cite{2019GCN.24182....1H};
$^{19}$\cite{2019GCN.24315....1X};
$^{20}$\cite{2019GCN.24285....1L};
$^{21}$\cite{2019GCN.24270....1H};
$^{22}$\cite{lundquist_saguaro_non_detection_190425};
$^{23}$\cite{paterson2021_saguaro_190425};
$^{24}$\cite{2019GCN.24206....1B};
$^{25}$\cite{2019GCN.24194....1H};
$^{26}$\cite{2019GCN.24225....1H};
$^{27}$\cite{2019GCN.24307....1A};
$^{28}$\cite{2019GCN.24256....1H};
$^{29}$\cite{2019GCN.24190....1X};
$^{30}$\cite{2019GCN.24183....1I};
$^{31}$\cite{2019GCN.24198....1A};
$^{32}$\cite{2019GCN.24320....1A};
$^{33}$\cite{2019GCN.24343....1A};
$^{34}$\cite{2019GCN.24234....1S};
$^{35}$\cite{2019GCN.24202....1P};
$^{36}$\cite{2019GCN.24314....1P};
$^{37}$\cite{2019GCN.24188....1P};
$^{38}$\cite{2019GCN.24201....1B};
$^{39}$\cite{2019GCN.24304....1W};
$^{40}$\cite{2019GCN.24367....1V};
$^{41}$\cite{ddoti_observations}, R. L. Becerra, private communication;
$^{42}$\cite{2019GCN.24219....1M};
$^{43}$\cite{2019GCN.24328....1M};
$^{44}$\cite{2019GCN.24227....1B};
$^{45}$\cite{2019GCN.24267....1L};
$^{46}$\cite{2019GCN.24459....1K};
$^{47}$\cite{2019GCN.24337....1B};
$^{48}$\cite{2019GCN.24229....1S};
$^{49}$\cite{2019GCN.24319....1I};
$^{50}$\cite{2019GCN.24214....1C};
$^{51}$\cite{2019GCN.24324....1H};
$^{52}$\cite{2019GCN.24325....1C};
$^{53}$\cite{2019GCN.24334....1T}.}
\end{deluxetable*}

\begin{deluxetable*}{lcccccc} 
\label{tab:all_obs}
\tablecaption{List of the observations used in this work.}
\tablehead{
\colhead{Facility/Instrument} & 
\colhead{MJD} & 
\colhead{R.A. [deg]} & 
\colhead{Dec. [deg]} & 
\colhead{Band} & 
\colhead{$5\sigma$ Limiting Mag.} & 
\colhead{Source}
}
\startdata
ZTF & 58598.3460 & 180.0000 & 62.1500 & r & 20.46 & Treasure Map \\
ZTF & 58598.3465 & -168.6100 & 54.9500 & r & 20.95 & Treasure Map \\
ZTF & 58598.3470 & -165.0502 & 47.7500 & r & 21.05 & Treasure Map \\
ZTF & 58598.3474 & -167.7500 & 40.5500 & r & 20.80 & Treasure Map
\enddata
\tablecomments{This table is published in its entirety in the machine readable format. A portion is shown here for guidance regarding its form and content.}
\end{deluxetable*}

\section{Analysis and Results}
\label{sec:analysis}
In order to measure the efficiency of the search for the GW190425 EM counterpart, we first analyze the observability of the GW localization region at the time of the merger (Section \ref{subsec:sky_conditions}), followed by the amount of the GW localization that was covered as a function of time, in terms of the total probability, area, and galaxy luminosity (Sections \ref{subsec:area_and_probability_coverage} and \ref{subsec:galaxy_luminosity_coverage}). We then examine the reported non-detection limiting magnitudes and compare them to light curves of the GW170817 kilonova and of kilonova models tuned to the parameters of GW190425 (Section \ref{subsec:limiting_magnitudes}). 

GW localizations are provided as Hierarchical Equal Area isoLatitude Pixelation\footnote{\url{https://HEALPix.sourceforge.io/}} \citep[HEALPix;][]{healpix} maps. These maps divide the sky into pixels of equal area and assign a probability value for the location of the GW source to each pixel. Each pixel is also assigned a distance to the source (assuming it is at that position) and a distance error estimate.
The number of pixels $N_{pixels}$ in the map is $12\cdot(nside)^2$ with $nside=2^n$ for some integer $n\geq0$. 
The localization maps provided for GW190425 have a resolution of $nside=256$, or $N_{pixels}=786,432$. This translates into pixels with an area of 0.052 deg$^{2}$ each. We perform all of our analysis using the maps in this resolution, as they were provided, except for the observability analysis (Section \ref{subsec:sky_conditions}), for which we downsample the maps to $nside=32$ (i.e. $N_{pixels}=12,288$ and a pixel area of 3.36 deg$^{2}$) for computational ease.

\subsection{Observability}
\label{subsec:sky_conditions}
We define that a region on the sky is ``visible'' at a point in time from a certain location on Earth when the Sun is at least \ensuremath{12^{\circ}} below the horizon, the Moon separation is at least \ensuremath{20^{\circ}} (the moon illumination was 66\% at the time of the GW190425 merger) and the airmass is less than 2.5. According to this definition, of the 9881 deg$^{2}$ (2400  deg$^{2}$) of the 90\% (50\%) final localization, 7783 deg$^{2}$ (2182 deg$^{2}$), or roughly 79\% (91\%), were visible during the 24 hours following the merger from the combined locations of all ground-based observatories listed in Table \ref{tab:observations_summary}.

\begin{figure} 
    \centering
    \includegraphics[width=0.5\textwidth]{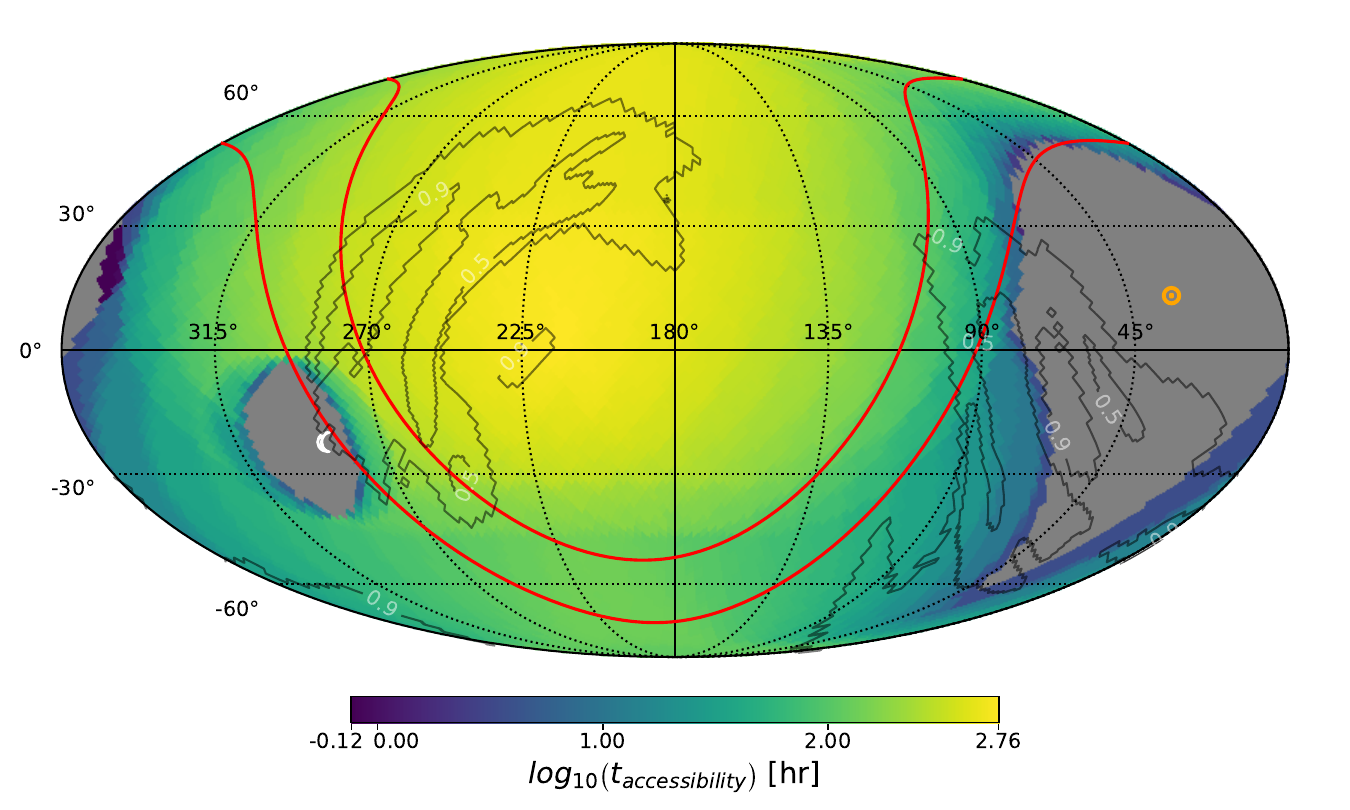}
    \caption{Accessible hours during the first day after the merger summed over all ground-based observatories listed in Table \ref{tab:observations_summary}. An accessible hour is one that have airmass $<2.5$ and not sun or moon-constrained, normalized to the FOV of each instrument (see text for details). Grey regions indicate inaccessible areas $(t_{accessibility}<5$ min). Black contours are of the GWTC-2 localization. Red contours denote a $\pm10^{\circ}$ band around the galactic plane. The sun and moon locations in the figure are at the time of the merger.}
    \label{fig:visibility_map}
\end{figure}

\begin{figure} 
    \centering
    \includegraphics[width=0.5\textwidth]{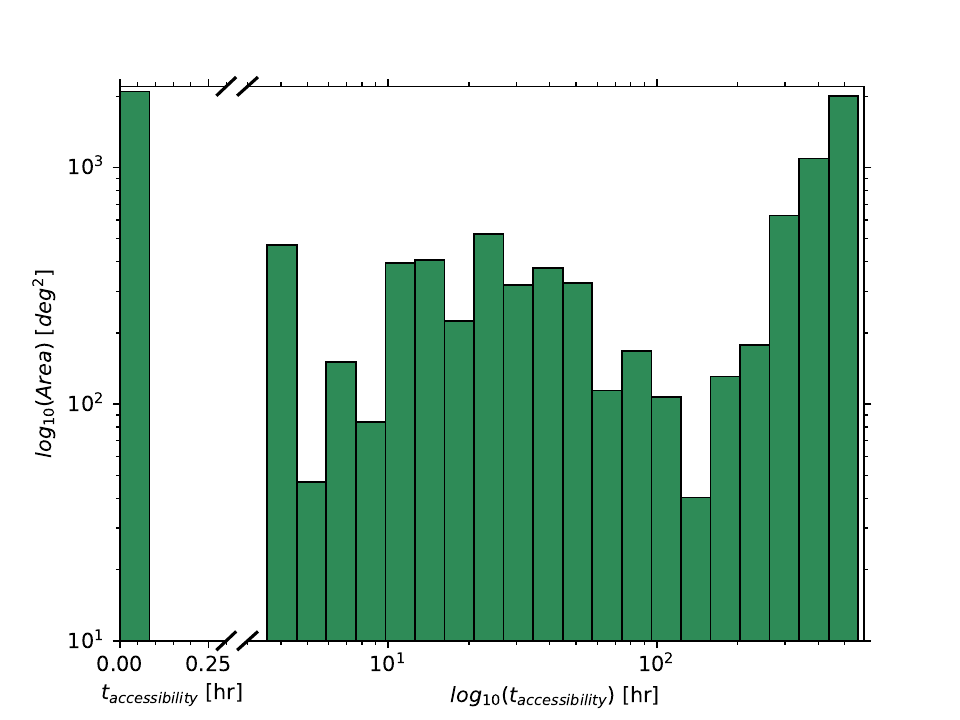}
    \caption{Accessible hours of the GWTC-2 90\% localization during the first day after merger. Out of the 90\% localization area, 2098 deg$^{2}$ were inaccessible ($t_{accessibility}<5$ min, leftmost bin), leaving 75\% of the probability accessible, with a median total accessibility of 41.5 hours.}
    \label{fig:visibility_histogram}
\end{figure}

Next, we define the ``accessibility'' of each area of the sky to each instrument as the amount of time per day that area is visible to that instrument, $t_{visible}$, weighted by the instrument FOV (to take into account that larger FOV instruments can cover more area simultaneously) and divided by the pixel area of the map (3.36 deg$^{2}$ in this case):
\begin{equation}
\label{eq:visible_time}
t_{accessibility}=t_{visible}\left(\frac{FOV}{3.36deg^{2}}\right)
\end{equation}
We define an inaccessible area as one with total accessibility (i.e. summed over all ground-based instruments considered here) $t_{accessibility}<5\ min$ (corresponding to a typical single exposure time\footnote{This is a conservative estimate, as ZTF and Pan-STARRS reported consecutive observations as fast as 38 and 54 seconds apart, respectively. Therefore, the true observability is likely even higher than what we present here.}). Using this definition, 6284 deg$^{2}$ of the sky were inaccessible due to sun and moon constraints, of which 2098 deg$^{2}$ were in the GWTC-2 90\% localization region. The rest of the localization region, encompassing 75\% of the probability, was visible with a median total accessibility time of 41.5 hours.
The total accessibility of the 90\% localization region during the first day since merger is presented in Figures \ref{fig:visibility_map} and \ref{fig:visibility_histogram}, and in Table \ref{tab:accessibility_summary}. This definition of accessibility is intended to provide a rough estimate of what was observable to the community. It does not take into account telescope-specific overheads, limitations, and observing strategies.

\begin{deluxetable}{lcccc} 
\label{tab:accessibility_summary}
\tablecaption{GW190425 accessible localization areas and probabilities, and median accessibility.}
\tablehead{
\colhead{Epoch} & \colhead{90\% Area} & \colhead{Accessible} & \colhead{Accessible} & \colhead{Median}
\\[-0.2cm]
\colhead{} & \colhead{} & \colhead{90\% Area} & \colhead{Probability} & \colhead{Accessibility}
\\[-0.2cm]
\colhead{} & \colhead{[deg$^{2}$]} & \colhead{[deg$^{2}$]} & \colhead{} & \colhead{[Hr]}
}
\startdata
Initial & 10,183 & 8216 & 72\% & 80.0 \\ \hline 
Update  & 7461 & 5833 & 76\% & 32.0 \\ \hline 
GWTC-2  & 9881 & 7783 & 75\% & 41.5
\enddata
\tablecomments{Median accessibility hours being greater than 24 means that certain areas could have been covered more than once during the first day, in total, by all instruments involved in the search.}
\end{deluxetable}

\subsection{Area and Probability Coverage}
\label{subsec:area_and_probability_coverage}

\begin{figure*} 
    \centering
    \includegraphics[width=\textwidth]{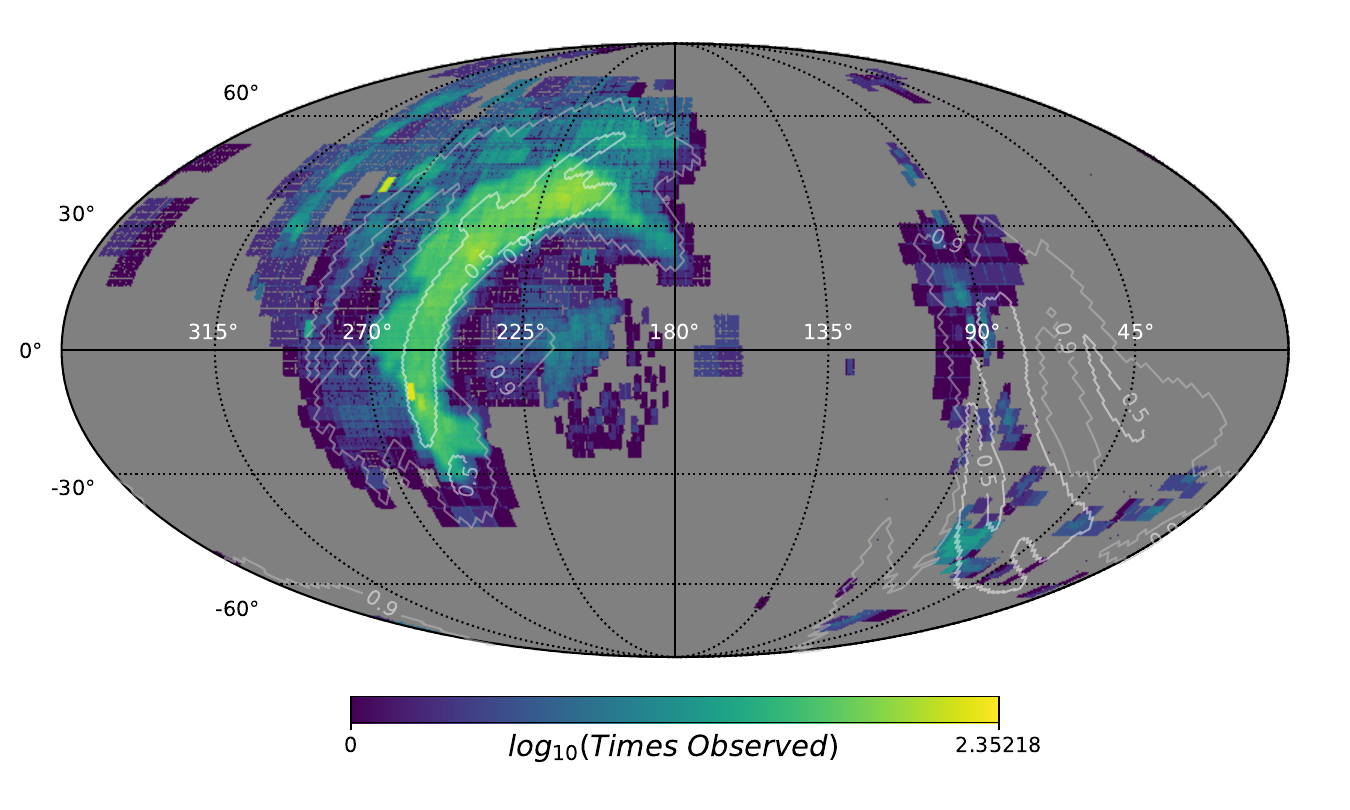}
    \caption{Total number of infrared, optical and ultraviolet followup observations up to 5.65 days post merger for GW190425. Grey areas are where no observations were reported. Lines mark the 50\% and 90\% probability regions of the final GWTC-2 localization. While some parts of the localization were observed over 100 times, some were not observed at all, even though they were accessible (Figure \ref{fig:visibility_map}).}
    \label{fig:observations_map}
\end{figure*}

We count the sky area and the probability observed up to 5.65 days after the time of the merger, both with and without overlap between observations (herafter denoted ``total coverage'' and ``unique coverage''), in order to measure how much of the localization was covered and how much of it was observed multiple times. Measuring total area and probability on a HEALPix map requires counting the pixels that overlap with an observed footprint. Due to the resolution of the map, the minimal area that can be counted per footprint is 0.052 deg$^{2}$. However, there are footprints with a smaller FOV (as seen in Table \ref{tab:observations_summary}). Here we use the {\tt{healpy}}\footnote{\url{https://healpy.readthedocs.io/}} python package, which includes two methods for counting (``querying'') pixels inside an observed footprint. The ``inclusive'' method counts every pixel that overlaps with the footprint, while the ``exclusive'' method only counts pixels that have their center fall inside the footprint. An illustration of these two methods is presented in Figure \ref{fig:inclusive_exclusive} in Appendix \ref{appendix:exclusive_inclusive}. We choose the exclusive method for our analysis, as it is a more conservative accounting of the actual area and probability covered, while the inclusive method over counts pixels, causing an overestimation of the area and probability covered.

Figure \ref{fig:observations_map} shows the total coverage of all reported followup observations up to 5.65 days post-merger outlined in Section \ref{sec:data}. Some regions, in the northern part of the localization, were observed more than 100 times while most of the accessible southern part of the localization was not observed at all (in Appendix \ref{appendix:north_south_analysis} we show that this is not entirely explained by the distribution of observing facilities, and hence is likely due to lack of coordination). 

The GW localization was updated 26 hours after the initial localization was released. At the time of the update, earlier footprints had their probability coverage changed. We take that into account by retroactively recounting the probability covered by all of the observations using the updated localization from the time it was released.
The area covered does not depend on the localization map and thus, no recounting of the area covered is needed at the time of the update. 
The sum of area and probability covered is performed in two different ways, first without regard to whether observations overlap or not, and second with overlapping regions counted only once.  We compare the probability covered to the 100\% probability of the entire sky and the area covered to the initial and updated 90\% probability areas.

\begin{figure} 
    \centering
    \includegraphics[width=0.5\textwidth]{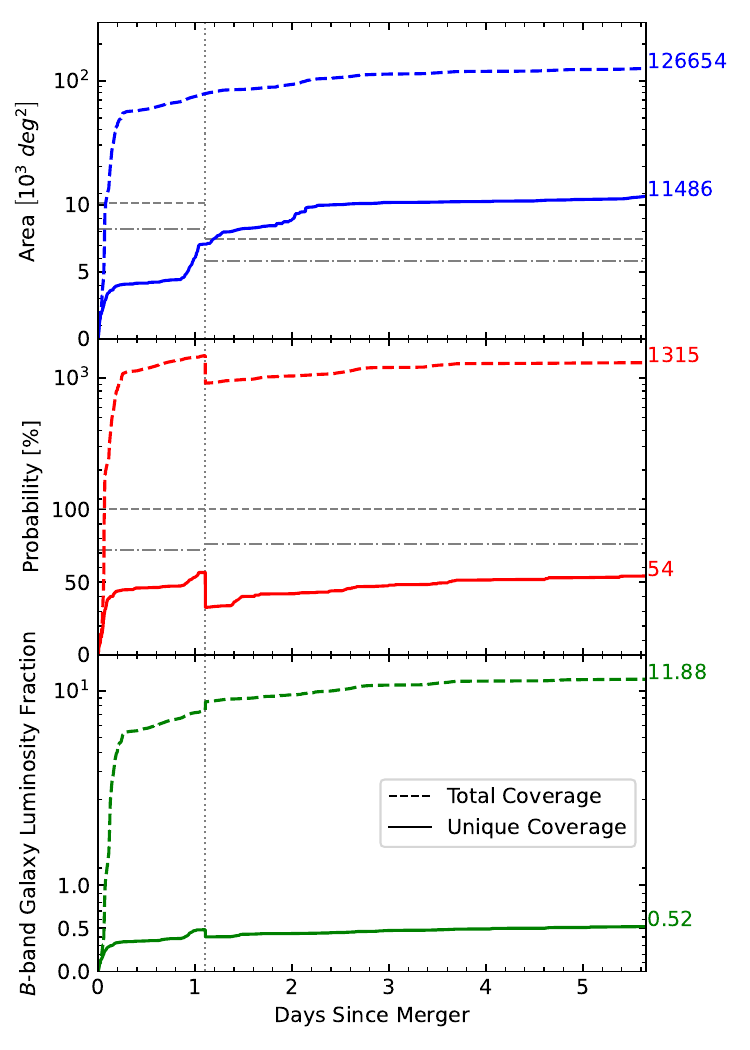}
    \caption{Cumulative area (top), probability (middle), and $B$-band galaxy luminosity fraction (bottom) observed during the 5.65 days following the GW190425 merger presented in symlog scale for the y-axis, with a linear range up to an area of 10,183 deg$^{2}$, probability of 100\% and luminosity fraction of 1, respectively. Colored dashed lines are for total coverage (including overlapping observations) while solid lines are for unique coverage (excluding overlapping observations). The observed $B$-band galaxy luminosity fraction is out of all galaxies with luminosity larger than $L^{*}_B$ (see Section \ref{subsec:galaxy_luminosity_coverage}) from the GLADE v2.3 catalog, that are within 3$\sigma$ of the mean estimated distance. Horizontal grey dotted lines denote the total area inside the 90\% localization region reported at the time and 100\% probability, for the top and middle plots, respectively, while the grey dash-dotted lines are for the accessible area and probability (see Section \ref{subsec:sky_conditions}). The vertical grey dotted line denotes the time of the updated localization. Coordinated followup, in principle, can cover a much larger area, probability and galaxy luminosity, significantly faster (i.e. on hour time scales instead of days) than the uncoordinated search.}
    \label{fig:area_prob_lum_calc}
\end{figure}

The area and probability covered as a function of time are shown in the top and middle panels of Figure \ref{fig:area_prob_lum_calc}, respectively. Within 5.65 days following the merger, 11486 deg$^{2}$ were uniquely covered. This is more than the area of the GWTC-2 90\% probability region, but not all of the area covered is in that region (Fig. \ref{fig:observations_map}). In fact, at 5.65 days post-merger, only 54\% of the probability (of the GWTC-2 localization) were uniquely covered, while, in total, enough observations were made to cover over 120,000 deg$^{2}$ by that time. According to Figure \ref{fig:area_prob_lum_calc}, a total area equivalent to the 90\% probability region was covered within three hours from merger. This means that \textit{in principle}, ideally coordinated observations could have covered the accessible 90\% localization on a time scale of hours from the field becoming observable\footnote{Weather and additional instrument-specific observability constraints could prolong this time estimate.}.

\subsection{Galaxy Luminosity Coverage}
\label{subsec:galaxy_luminosity_coverage}
We define the galaxy luminosity coverage as the fraction of $B$-band luminosity of galaxies observed out of the cumulative $B$-band luminosity of all galaxies with $L>L_{B}^{*}$ in the 90\% localization volume within 3$\sigma$ of the mean localization distance. Here, $L_{B}^{*}$ is the $B$-band luminosity corresponding to $L^{*}$, the characteristic luminosity of the Schechter Function \citep{schechter_function}. We adopt a corresponding absolute magnitude in the $B$ band of $M_B^{*}=-20.7$, following \cite{o2_summary}. The GLADE catalog is less complete to galaxies fainter than $L^{*}$ at large distances \citep[see][for more details]{dalya_glade2p3}.

We count only galaxies that have a $B$-band apparent magnitude and a distance estimate in Version 2.3 of the GLADE catalog. There are a total of 191,033 galaxies (133,240) inside the initial (updated) 90\% localization. Of these 47,664 (34,302) have $L>L^{*}_{B}$, a luminosity fraction of 0.72 (0.72).

We count galaxies that are inside observed footprints and the localization volume. The bottom panel of Figure \ref{fig:area_prob_lum_calc} shows the galaxy luminosity coverage as a function of time. After 5.65 days from the merger, only 52\% of the $B$-band luminosity in galaxies inside the updated localization region was uniquely observed, while 11.88 times the localization luminosity was observed in total due to overlapping observations.

\subsection{Kilonova Detection}
\label{subsec:limiting_magnitudes}

\begin{deluxetable*}{llll}
\label{tab:mosfit_priors}
\tablecaption{Parameters and distributions used for generating the ensemble of light curves from the \cite{nicholl_models} BNS merger kilonova model.}
\tablewidth{0pt}
\tablehead{\colhead{Parameter} & \colhead{Description} & \colhead{Type} & \colhead{Values}}

\startdata
$\chirpmass^a\ [M_{\odot}]$ & Chirp mass & Gaussian & $\mu=1.44$, $\sigma=0.02$ \\
$q^a$ & Mass ratio (high spin case) & uniform & $\min=0.4$, $\max=1.0$ \\
 & Mass ratio (low spin case) & uniform & $\min=0.8$, $\max=1.0$ \\
s & Shock density profile power law index & fixed & 1.0 \\
$M_{TOV}\ [M_{\odot}]$ & NS maximum theoretical mass & fixed & 2.2 \\
$\alpha^b$ & Enhancement of blue ejecta & fixed & 1.0 \\
$disk_{frac}^b$ & Fraction of disk ejected & fixed & 0.15 \\
$\cos\theta^b$ & Cosine of viewing angle & uniform & $\min=0.0$, $\max=1.0$ \\
$\cos\theta_{cocoon}^b$ & Cosine of cocoon opening angle & uniform & $\min=0.707$, $\max=1.0$ \\
$\cos\theta_{open}^b$ & Cosine of squeezed ejecta opening angle & uniform & $\min=0.707$, $\max=1.0$ \\
$\log N_H^b$ & Hydrogen column density & uniform & $\min=19$, $\max=23$ \\
\enddata
\tablenotetext{a}{Based on the inferred values in \cite{gw190425} for the cases of high and Low spin.}
\tablenotetext{b}{Based on the priors used in \cite{gw170817constraints}.}
\end{deluxetable*}

Even if the localization would have been covered as efficiently as possible, there is still a possibility that the EM counterpart was too faint to have been detected. In order to check this, we compare the non-detection limiting magnitudes reported for each of the followup observations to the light curve of GW170817 and to model kilonova light curves for GW190425. 

There is a large uncertainty in predicting the light curve of a kilonova based on the GW parameters, given that only one such event was observed. Therefore, the results of this analysis depend on the chosen model and its parameters. Here we follow \cite{foley_updated_parameters} and \cite{nicholl_models} who predicted the GW190425 kilonova using the \cite{kasen_models} and \cite{nicholl_models} models respectively. 

Three additional models are explored in Appendix \ref{appendix:additional_models}: a ``worst-case'' mass ratio and cocoon geometry for the \cite{nicholl_models} model with no blue component at all, a spherical two ejecta component model with high variance in the component angular geometry by \cite{bulla2019_possis}, and a two-ejecta component model with varying morphology and $Y_{e}$ content by \cite{wollaeger2021_supernu}.

We use the InfraRed Survey Archive (IRSA) dust extinction service queries\footnote{\url{https://astroquery.readthedocs.io/en/latest/ipac/irsa/irsa_dust/irsa_dust.html}} to correct all reported magnitude limits for Milky Way dust extinction at the center of their footprint by using the \cite{irsa} galactic dust extinction estimates. For unfiltered observations, we assume $r$-band extinction. We then translate all limits to absolute magnitudes using the mean distance of the pixel that overlaps with the footprint center.

The GW170817 data is taken from the \cite{Arcavi_2018} compilation of observations by \cite{andreoni_gw170817_obs, arcavi_gw170817_obs, coulter_gw170817_obs, cowperthwaite_gw170817_obs, diaz_gw170817_obs, drout_gw170817_obs, evans_gw170817_obs, hu_gw170817_obs, kasliwal_gw170817_obs, lipunov_gw170817_obs, pian_gw170817_obs, shapee_gw170817_obs, smartt_gw170817_obs, tanvir_gw170817_obs, troja_gw170817_obs, utsumi_gw170817_obs, valenti_gw170817_obs} and \cite{pozanenko_gw170817_obs}. 

\begin{deluxetable*}{cccc}
\tablecaption{Ejecta parameters for two \cite{kasen_models} models with lanthanide fractions $X_{lan}=-2$ and $X_{lan}=-4$, and two \cite{nicholl_models} models (low and high spin cases) with the priors from Table \ref{tab:mosfit_priors}.}
\tablewidth{0pt}
\tablehead{\colhead{Model} & \colhead{$M_{ej}$ [M$_{\odot}$]} & \colhead{$v_{k}\ [c]$} & \colhead{$\kappa_{ej}$ [cm$^{2}$/g]}}

\startdata
\cite{kasen_models}$^{a}$ & & & \\
$X_{lan}=-2$ & $0.04$ & $0.15$ & $1$ \\
$X_{lan}=-4$ & $0.04$ & $0.10$ & $10$ \\ \hline
\cite{nicholl_models} High-spin$^{b}$ & & & \\
Blue Component & $0.00065\pm0.00013$ & $0.1550\pm0.0022$ & $0.5$ \\
Red Component & $0.0408\pm0.0044$ & $0.552\pm0.0019$ & $10$ \\
Purple Component & $0.00565\pm0.00058$ & $0.0505\pm0.0010$ & $\simeq5.6$ \\
Total/Mean & $0.0471\pm0.0040$ & $0.1973\pm0.0068$ & $\simeq8.4$ \\ \hline
\cite{nicholl_models} Low-spin$^{b}$ & & & \\
Blue Component & $0.00223\pm0.00013$ & $0.1666\pm0.0012$ & $0.5$ \\
Red Component & $0.00401\pm0.00028$ & $0.2420\pm0.0007$ & $10$ \\
Purple Component & $0.01120\pm0.00051$ & $0.0578\pm0.0003$ & $\simeq5.55$ \\
Total/Mean & $0.01740\pm0.00053$ & $0.1186\pm0.0023$ & $\simeq5.87$ \\
\enddata
\tablenotetext{a}{Taken from \url{https://github.com/dnkasen/Kasen_Kilonova_Models_2017} and \cite{kasen_models}.}
\tablenotetext{b}{Calculated from binary parameters based on \url{https://github.com/mnicholl/kn-models-nicholl2021}.}
\label{tab:model_results}
\end{deluxetable*}

The \cite{kasen_models} model is based on a 1D Monte Carlo simulation of the radiative transfer of photons from the radioactive decay of $r$-process elements through kilonova ejecta. The ejecta is modeled using three parameters: ejecta mass $M_{ej}$, characteristic expansion velocity $v_{k}$, and lanthanide fraction power index $X_{lan}$ (where the lanthanide fraction is $10^{X_{lan}}$). The calculation produces a spectral energy distribution of the emission. We use {\tt{pyphot}}\footnote{\url{https://mfouesneau.github.io/pyphot/index.html}} to create synthetic light curves in the relevant bands from these model spectra. We use the ejecta parameters assumed by \cite{foley_updated_parameters} for the GW190425 kilonova: $M_{ej}=0.04$ M$_{\odot}$, $v_{k}=0.15c$ and $X_{lan}=-2$, which they propose assuming a BNS system with masses 1.4 M$_{\odot}$ and 2.0 M$_{\odot}$. This lanthanide fraction results in a high opacity of ${\kappa_{ej}}\approx10$ cm$^{2}$/g and a transient that is observable mostly in the infrared. This led \cite{foley_updated_parameters} to conclude that the GW190425 kilonova would not have likely been observable to most facilities that participated in the search. To illustrate the effect of the assumed lanthanide fraction we also use a model with $M_{ej}=0.04$ M$_{\odot}$, $v_{k}=0.10c$ and $X_{lan}=-4$. The lower lanthanide fraction compared to the \cite{foley_updated_parameters} parameters results in a lower opacity of ${\kappa_{ej}}\approx1$ cm$^{2}$/g, leading to a bluer and brighter kilonova.

\cite{nicholl_models} provide an analytic 2D kilonova model that assumes a three-component ejecta: one component is the ``tidal ejecta'' in the equatorial region and a second is the ``polar ejecta'' concentrated in the polar region. For both components, a blackbody is assumed. The third component is from GRB-shocked material in a ``cocoon'' around the polar region \citep{Nakar_grb_cocoon}. The model parameters relate directly to the GW-measured binary system chirp mass $\mathcal{M}$ and mass ratio $q$, and to neutron star (NS) equation of state parameters, such as the NS maximum theoretical mass $M_{TOV}$. These are used to calculate the mass $M_{ej}$, velocity $v_{k}$, and the opacity ${\kappa_{ej}}$ of each ejecta component corresponding to three opacity regimes: the ``red'' high opacity (low $Y_{e}$) tidal ejecta, the  ``blue''  low opacity (high $Y_{e}$) polar ejecta, and the intermediate-opacity ``purple'' ejecta. The blue and red opacities are assumed to be $\kappa_{ej,blue}=0.5$ cm$^{2}$/g and $\kappa_{ej,red}=10$ cm$^{2}$/g, respectively. The asymmetry of the ejecta is parameterized by the opening angle around the poles ${\theta_{open}}$, which separates the polar ejecta and the tidal ejecta, and the opening angle of the shocked cocoon ${\theta_{cocoon}}$. Finally, the observer's viewing angle ${\theta}$ and the host hydrogen column density $N_H$ are taken as parameters as well. The model is implemented in the Modular Open Source Fitter for Transients ({\tt{MOSFiT}}; \citealt{mosfit}), which can be used to generate light curves in various bands. We generate an ensemble of models, using the binary parameters $\mathcal{M}$ and $q$ from \cite{gw190425} inferred for two cases based on the spin of the neutron stars prior to merger: (1) low spin ($\chi<0.05$, where $\chi={cS}$/${\left(Gm^2\right)}$ is the dimensionless spin parameter, with $c$ the speed of light, $S$ the spin angular momentum magnitude and $m$ the mass, for each component), and (2) allowing for high spin ($\chi<0.89$). The enhancement of blue ejecta due to magnetically driven winds is set to $\alpha=1.0$ (no enhancement), as the remnant mass is expected to lead to a prompt collapse \citep{nicholl_kn_mosfit}. For the rest of the parameters, we sample 100 realizations of the ensemble (for each of the low and high spin cases) from the prior distributions that \cite{nicholl_models} used for their GW170817 kilonova fit (which included three fixed parameters, as we repeat here). The full set of parameters and the values used are described in Table \ref{tab:mosfit_priors}. The Ejecta parameters for all models are summarized in Table \ref{tab:model_results}.

\begin{figure*} 
\includegraphics[width=\textwidth]{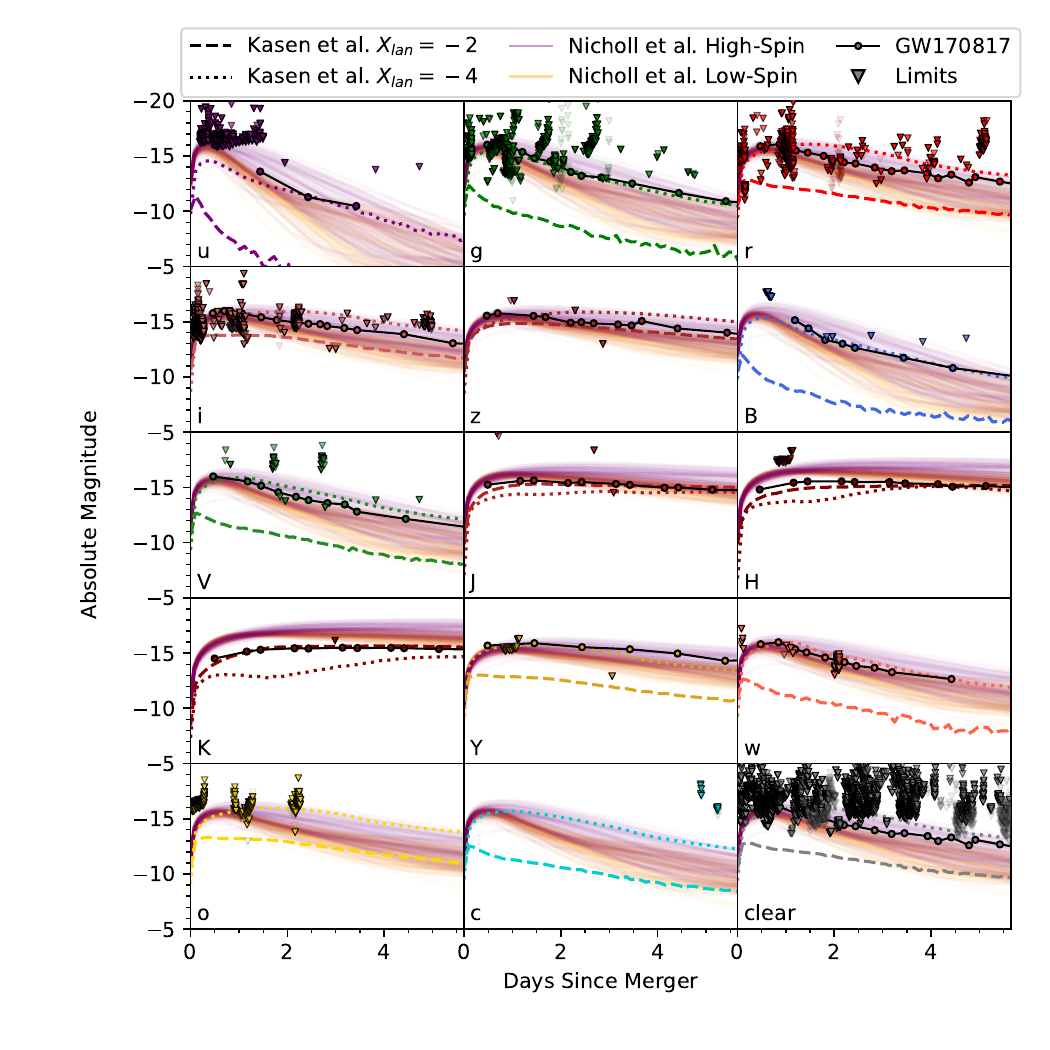}
\caption{5$\sigma$ non-detection limits of the GW190425 kilonova (triangles) compared to two kilonova models from \cite{kasen_models} with dashed and dotted lines for $X_{lan}=-2$ and $X_{lan}=-4$, respectively, two ensembles of models from \cite{nicholl_models} with semi-transparent solid purple and orange lines for high and low spin cases, respectively, and the GW170817 kilonova (circles; see text for data sources). The upper limit triangles are displayed fully opaque for observations within the 50\% GWTC-2 
localization region, partially opaque for observations outside the 50\% GWTC-2 
localization region but within the 90\% GWTC-2 localization region, and almost transparent for observations outside the 90\% GWTC-2 localization region. The limits are converted to absolute magnitude after Milky Way extinction correction and using the distance estimate at the HEALPix closest to the center of the pointing for each limit. Data in the `clear' band is unfiltered and is compared to models in the $r$-band. Limits with magnitudes brighter than -20 are not shown.}
\label{fig:magnitude plots}
\end{figure*}

The comparison of reported non-detection limiting magnitudes to the GW170817 kilonova and the model light curves described above can be seen in Figure \ref{fig:magnitude plots}. \cite{smartt_ps_atlas} split the Pan-STARRS pointings into 6209 ``skycells'' (see \citealt{pan_starrs} for more details), each with a footprint of $0.16\ deg^{2}$, and calculated limiting magnitudes per skycell (except for the $z$-band observations, for which no limiting magnitudes were calculated). Here we plot each of the 6147 $i$- and $w$-band skycell limits included in the 232 Pan-STARRS pointings.

We find that, had the GW190425 kilonova been similar to that of GW170817, it could have been detected by some of the instruments despite the larger distance of GW190425 compared to that of GW170817.

\begin{figure}[h] 
    \centering
    \includegraphics[width=\linewidth]{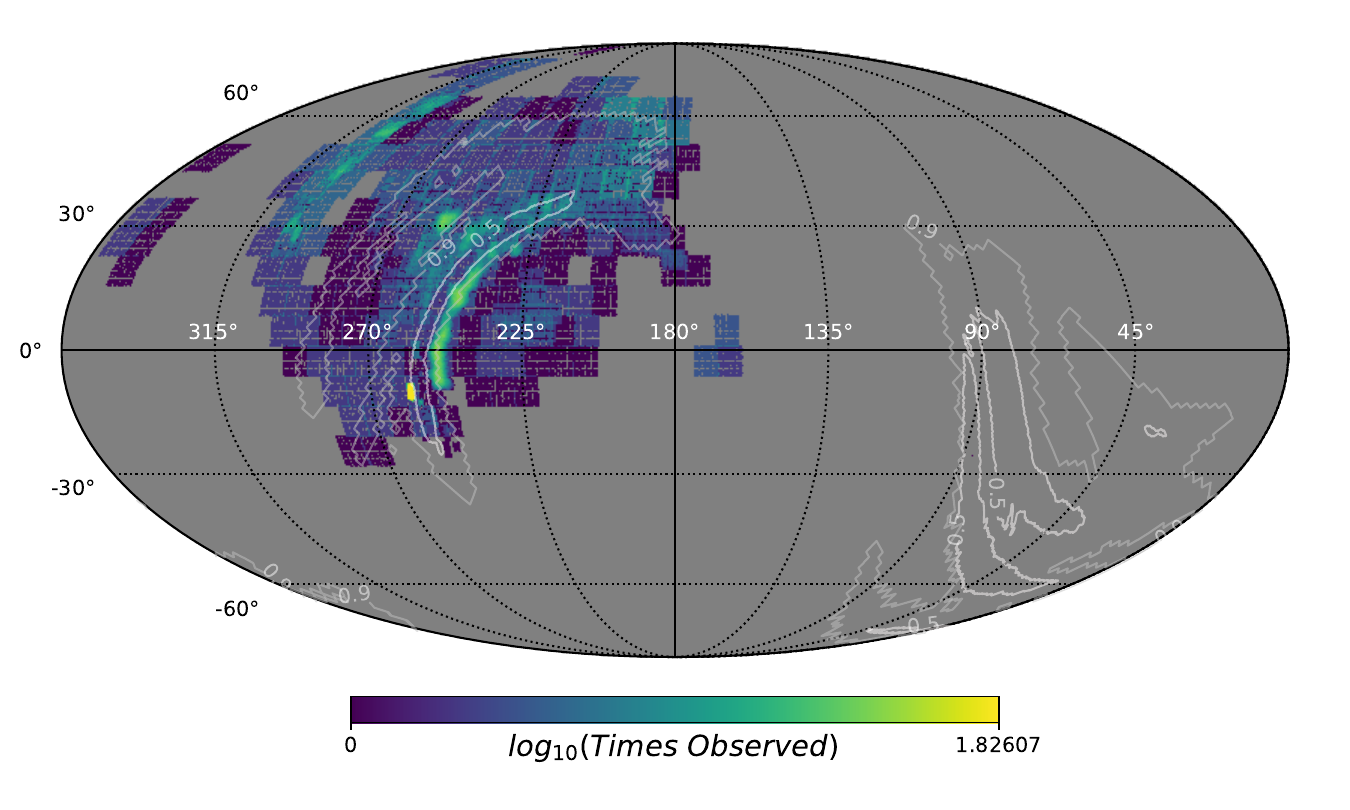}
    \caption{Same as Figure \ref{fig:observations_map}, using only observations that were deep enough to detect a GW170817-like kilonova at the location of GW190425.}
    \label{fig:coverage_map_under_gw170817}
\end{figure}

\begin{figure}[h] 
    \centering
    \includegraphics[width=\linewidth]{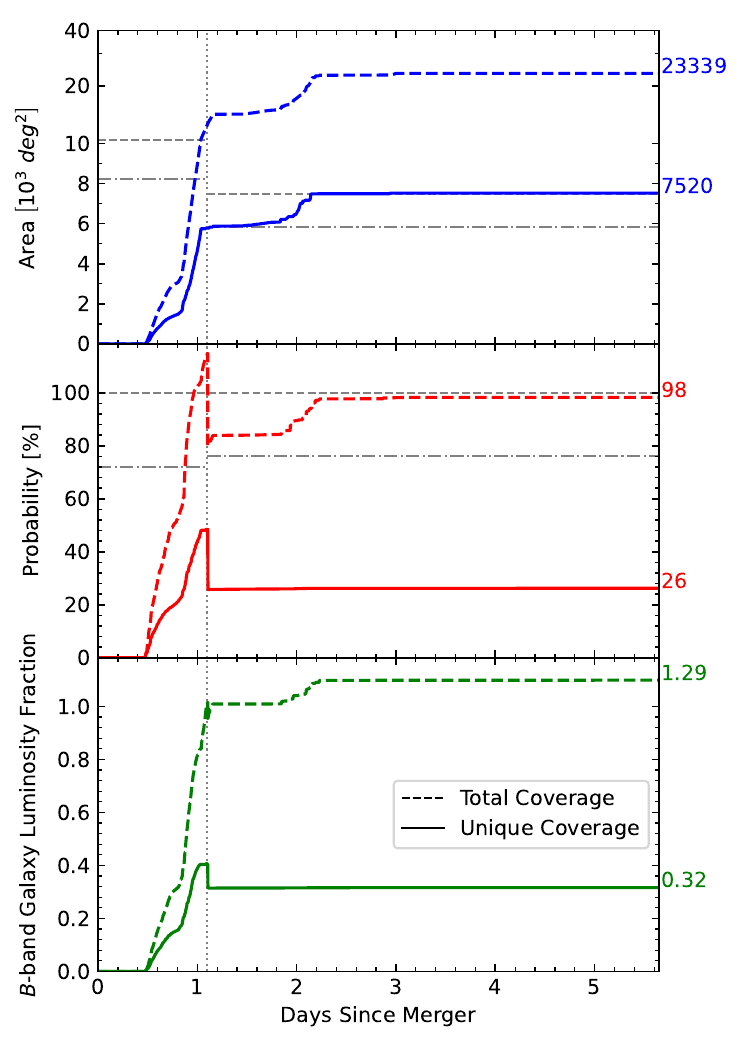}
    \caption{Same as Figure \ref{fig:area_prob_lum_calc}, using only observations that were deep enough to detect a GW170817-like kilonova at the location of GW190425. Since there are no data for the GW170817 kilonova prior to 0.45 days after merger, we do not consider observations obtained for GW190425 within that time range.}
    \label{fig:coverage_under_gw170817}
\end{figure}

We repeat the analysis of Sections \ref{subsec:area_and_probability_coverage} and \ref{subsec:galaxy_luminosity_coverage} using only observations that were deep enough to detect a GW170817-like kilonova. We select these observations as having limits deeper than the linearly interpolated light curve of the GW170817 kilonova (scaled to the distance of GW190425). Here we ignore $c$- and $o$-band observations, compare clear limits to the $r$-band light curve of the GW170817 kilonova, and ignore all observations obtained before 0.45 days post merger, as there are no GW170817 kilonova observations at those times\footnote{In practice, some of the observations obtained for GW190425 in this time span could likely have detected a GW170817-like kilonova here, therefore our results are conservative.}. Our results are presented in Figures \ref{fig:coverage_map_under_gw170817} and \ref{fig:coverage_under_gw170817}. There are 1358 observations sensitive enough to detect a GW170817-like kilonova at the position of GW190425. These observations could have covered an area of approximately 23300 deg$^{2}$, 98\% of the probability, and 1.3 times the $B$-band galaxy luminosity, in total. Due to overlaps, the unique area covered by these observations was about 7500 deg$^{2}$, the  unique probability reached was 26\%, and the unique galaxy luminosity fraction obtained was 0.32.

Given the different merger parameters of GW190425 compared to those GW170817, it is possible that the GW190425 kilonova was fainter than that of GW170817, though estimates are highly model dependent. Indeed, the high-lanthanide \cite{kasen_models} model for the GW190425 kilonova predicts much fainter optical emission compared to the GW170817 kilonova. According to that model, the kilonova would just have barely been detected in a few of the bands used in the search. However, the low-lanthanide scenario predicts a kilonova similar to (and in the $r$, $i$ and $z$ bands even brighter than) GW170817, which could have been detected with the observations obtained. The GW170817 kilonova could only be explained with multiple ejecta components of different opacities. If the same is true for GW190425, then the emission from its kilonova might even be as luminous as the sum of these two \cite{kasen_models} models.

The \cite{nicholl_models} models, which already take into account multiple ejecta components, predict a kilonova that could have been detected in the optical bands. This is true for both the high-spin and low-spin assumptions.

\section{Discussion and Conclusions}
\label{sec:conclusions}

We analyzed all reported ultraviolet, optical and infrared followup observations of GW190425, the only robust BNS merger detected since GW170817, and showed that:
\begin{enumerate}
    \item Roughly 75\% of the localization region of GW190425 was accessible at the time of the trigger due to sun and moon constraints.
    \item Enough observational resources were invested in the followup of GW190425 to allow the accessible part of the 90\% probability region to have been covered potentially in a few hours. Instead, several regions were observed over 100 times, while others were never observed. More than 5 days after the merger, only $\sim$50\% of the probability was covered, the majority of it in the northern hemisphere by both northern and southern facilities, despite the southern part of the localization being under observed.
    \item If the GW190425 kilonova were similar to the GW170817 kilonova (as some models suggest), it could have been securely detected. According to more conservative models without a blue emission component or having a less favorable ejecta geometry, the kilonova might have been at best only marginally detected around peak. 
\end{enumerate}

Our results only take into account observations undertaken by facilities which decided to followup GW190425. We do not have complete information on which additional facilities preferred not to trigger given the large localization region, but which might have participated in the search, had they known that roughly half of the probability was already being covered by others. In addition, we do not know which facilities could have focused on obtaining deeper limits, rather than covering a wider area, had they known about the existing coverage by other facilities in real time.

We conclude that even for relatively distant BNS mergers with large localizations, coordinated followup can significantly increase the chances of discovering an EM counterpart. Furthermore, when no counterparts are detected, a lack of coordination weakens our ability to constrain models of kilonova emission, given the amount of localization probability that was not covered, and for which no non-detection limits are available. For smaller localizations, coordination can still help find the kilonova sooner. Observations on few-hour time scales are critical for constraining emission models \citep[e.g.][]{Arcavi_2018}.

\cite{Ackley2020} list some of the candidate counterparts to the neutron-star - black-hole merger GW190814 and find little overlap in candidates between the different surveys, although they did overlap in the search area. This demonstrates that some overlap in observations is necessary (given different observing bands, depths and candidate search algorithms). However, such overlap should still be made as efficient as possible through coordination.

While a small fraction of the facilities, notably those with wide-field instruments, are responsible for a large fraction of the total coverage, coordination among these facilities is still important. In addition, smaller field-of-view instruments gain from knowing about locations, filters and depths covered by the larger facilities, to make their own strategy adjustments, which can allow for necessary overlap, gap-filling, and deeper coverage of certain areas. In fact, the GW190425 localization could have in principle been fully covered even without ZTF and Pan-STARRS, and a third of it covered by observations deep enough to see a GW170817-like kilonova not including ZTF and Pan-STARRS (more than twice what was actually covered by such observations without coordination; Appendix \ref{appendix:no_ztf_and_ps_analysis}). This demonstrates that our results are not driven by one or two dominant surveys.

Real-time coordination in such a competitive field, requiring rapid response, is challenging. Tools like the {\tm} have been built exactly to overcome this challenge. We encourage the community to report their pointings to the {\tm}, and to use the information there to guide their search, even if it means searching lower probability areas (instead of contributing the 100th observation to a higher probability region). So far, approximately 30,000 observations by various instruments have been reported to the {\tm} since its inception (both retroactively and in real time). Specifically, for the recent possible neutron-star -- black-hole merger S240422ed\footnote{\url{https://gracedb.ligo.org/superevents/S240422ed/view/}}, over 2200 observations were reported to the {\tm} in real time by 15 facilities. This demonstrates the desire of the community to coordinate followup in general, and their engagement with {\tm} in particular.

In addition to the {\tm}, several other tools have been developed in recent years to address different challenges in the follow-up process. \cite{coulter2024_teglon} show that {\tt{Teglon}} is able to reduce the localization region of GW190425 by $\sim$50\% by redistributing some of the probability to galaxies at the correct distance. Efficiency can be improved further by scheduling, tracking, and automatically reporting observations using various ``marshals'' or Target and Observations Managers \citep[TOMs;][]{street2024_tomtoolkit} such as {\tt SkyPortal} \citep{coughlin2023_skyportal} and {\tt YSE-PZ} \citep{coulter2023_ysepz}, and by communicating findings in a rapid and machine readable format, using, for example, {\tt{Hermes}} \citep{nation2024_hermes}. Such systems, ideally working seamlessly together, could further enhance the efficiency of follow-up efforts.

With the lack of additional significant BNS merger discoveries since GW190425, it is possible that the BNS merger rate is on the lower end of its large uncertainty range, and events as nearby and well localized as GW170817 are likely very rare. If we wish to unleash the potential of GW-EM multi-messenger astronomy for nuclear physics, astrophysics, and cosmology, we must better coordinate our GW followup searches.

~\\

We thank the anonymous referees for insightful and salient comments, as well as A. Tohuvavohu, M. Nicholl, J. Gillanders, A. Levan and M. Fraser for helpful discussions. We further thank A. Tohuvavohu for assisting with querying {\tm} and M. Nicholl for help using the kilonova model implemented in {\mosfit}.
I.A. acknowledges support from the European Research Council (ERC) under the European Union’s Horizon 2020 research and innovation program (grant agreement number 852097), and from the Pazy foundation (grant number 216312). I.K. and I.A. acknowledge support from the Israel Science Foundation (grant number 2752/19), from the United States - Israel Binational Science Foundation (BSF; grant number 2018166).
This paper has made use of version 2.3 of the GLADE galaxy catalog \citep{dalya_glade2p3}, the NASA/IPAC Infrared Science Archive, which is funded by the National Aeronautics and Space Administration and operated by the California Institute of Technology \citep{irsa}, the SIMBAD database,
operated at CDS, Strasbourg, France, and the SVO Filter Profile Service, supported by the Spanish MINECO, through grant AYA2017-84089 \citep{svo_rodrigo2012, svo_rodrigo2020}.
\software{{\qcr astropy} \citep{astropy2013, astropy2018, astropy2022}, {\qcr healpy} \citep{healpix}, {\qcr matplotlib} \citep{matplotlib}, {\qcr MOSFiT}, {\qcr numpy} \citep{numpy}, {\qcr pandas}, {\qcr pyphot}, {\qcr scipy} \citep{scipy_virtanen2020}.}

\bibliography{references}
\bibliographystyle{aasjournal}

\begin{appendix}
\section{Healpy Exclusive vs. Inclusive Methods}
\label{appendix:exclusive_inclusive} 
\begin{figure}[h] 
\includegraphics[width=\textwidth]{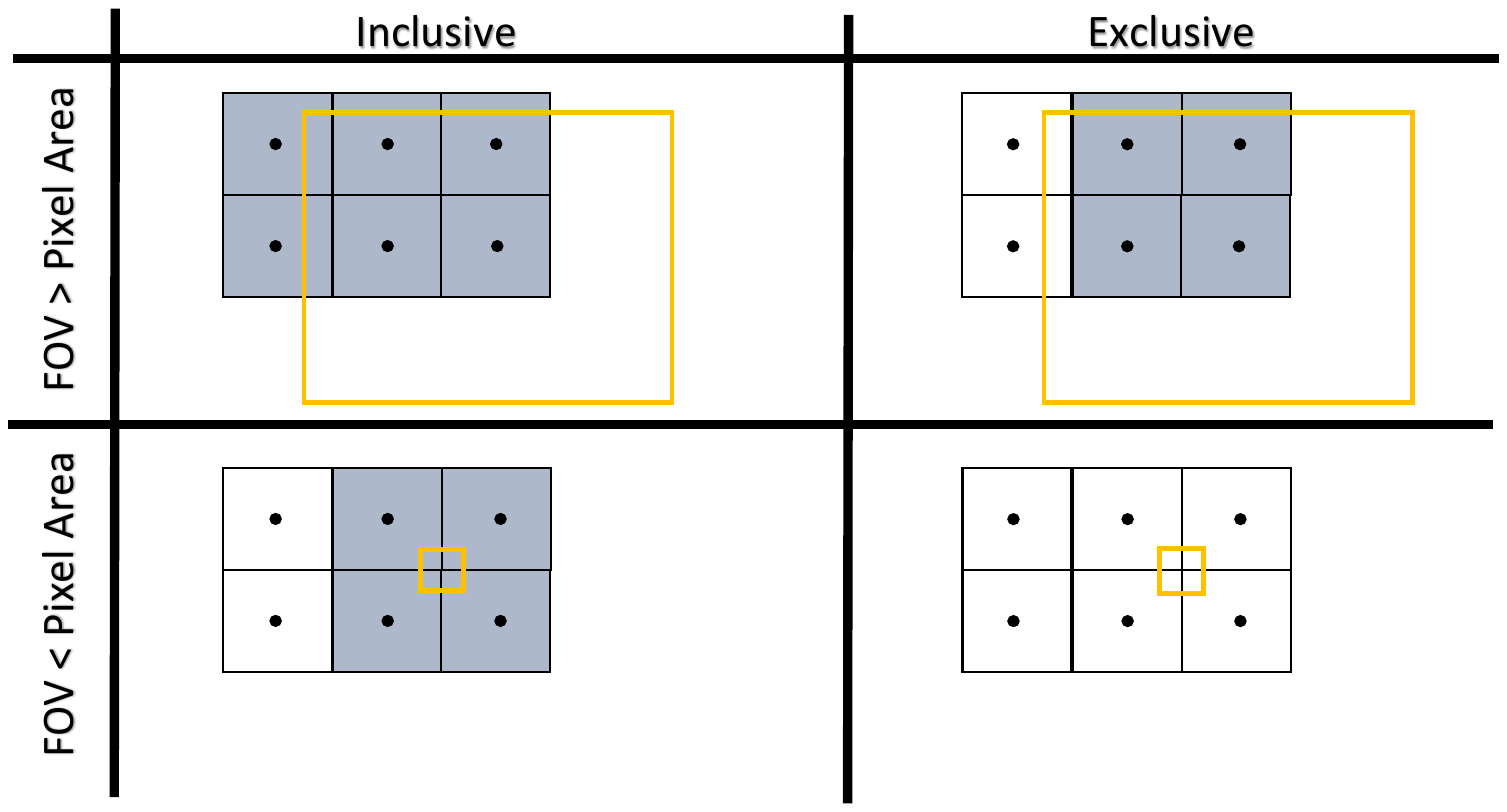}
\caption{\label{fig:inclusive_exclusive}
An illustration of the exclusive and inclusive {\healpy} pixel query methods. Each black square represents a pixel centered around a black dot and the yellow rectangles are hypothetical observational footprints. Grey pixels are the ones returned according to each method and white pixels are ignored. The inclusive method (left) returns every pixel that intersects with the observed footprint, while the exclusive method (right) returns only pixels with centers included in the observed footprint. For footprints with a field of view (FOV) much smaller than the pixel area (bottom), the inclusive method can significantly overestimate the area covered.}
\end{figure}

We present an illustration of the inclusive vs. the exclusive methods of {\healpy} pixel queries in Figure \ref{fig:inclusive_exclusive}. The inclusive method can significantly overestimate the area covered by a footprint with a field of view smaller than the pixel size. The exclusive method is much more conservative, and hence we chose it for our analysis. 

\section{Coverage Analysis by Hemisphere}
\label{appendix:north_south_analysis}

\begin{deluxetable}{lccccc}
\label{tab:visibility_by_hemishpere}
\tablecaption{GW190425 visible localization areas and probabilities per hemisphere.}
\tablehead{
\colhead{Hemisphere} & \colhead{Epoch} & \colhead{90\% Area} & \colhead{Visible 90\% Area} & \colhead{Probability} & \colhead{Visible Probability}
\\[-0.2cm]
\colhead{} & \colhead{} & \colhead{[deg$^2$]} & \colhead{[deg$^2$]} & \colhead{} & \colhead{}
}
\startdata
Northern & Initial & 4808 & 4521 & 53\% & 48\% \\
& Updated & 3084 & 2896 & 31\% & 26\% \\ \hline
Southern & Initial & 5374 & 3695 & 47\% & 24\% \\
& Updated & 4377 & 2937 & 69\% & 50\%
\enddata
\end{deluxetable}

\begin{figure}[h] 
    \centering
    \includegraphics[width=0.8\textwidth]{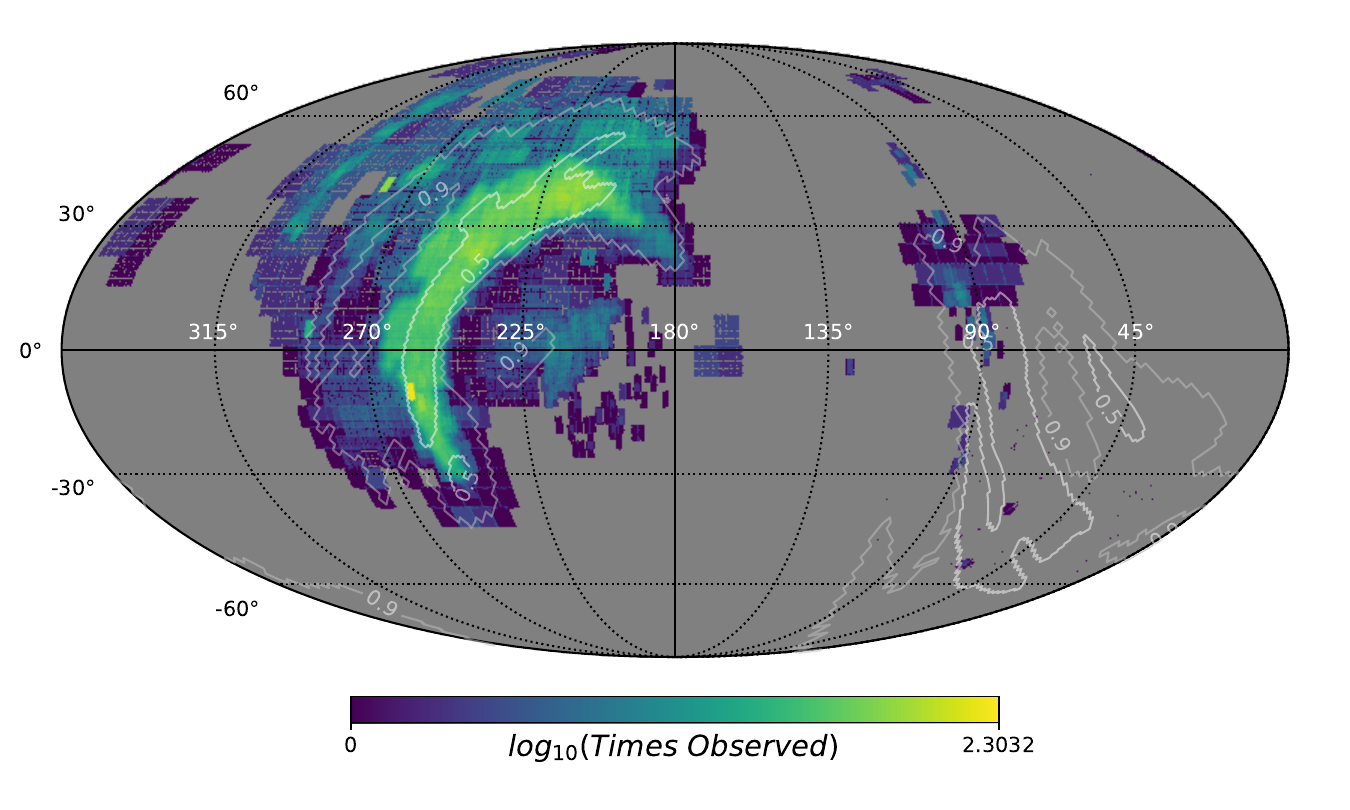}
    \includegraphics[width=0.8\textwidth]{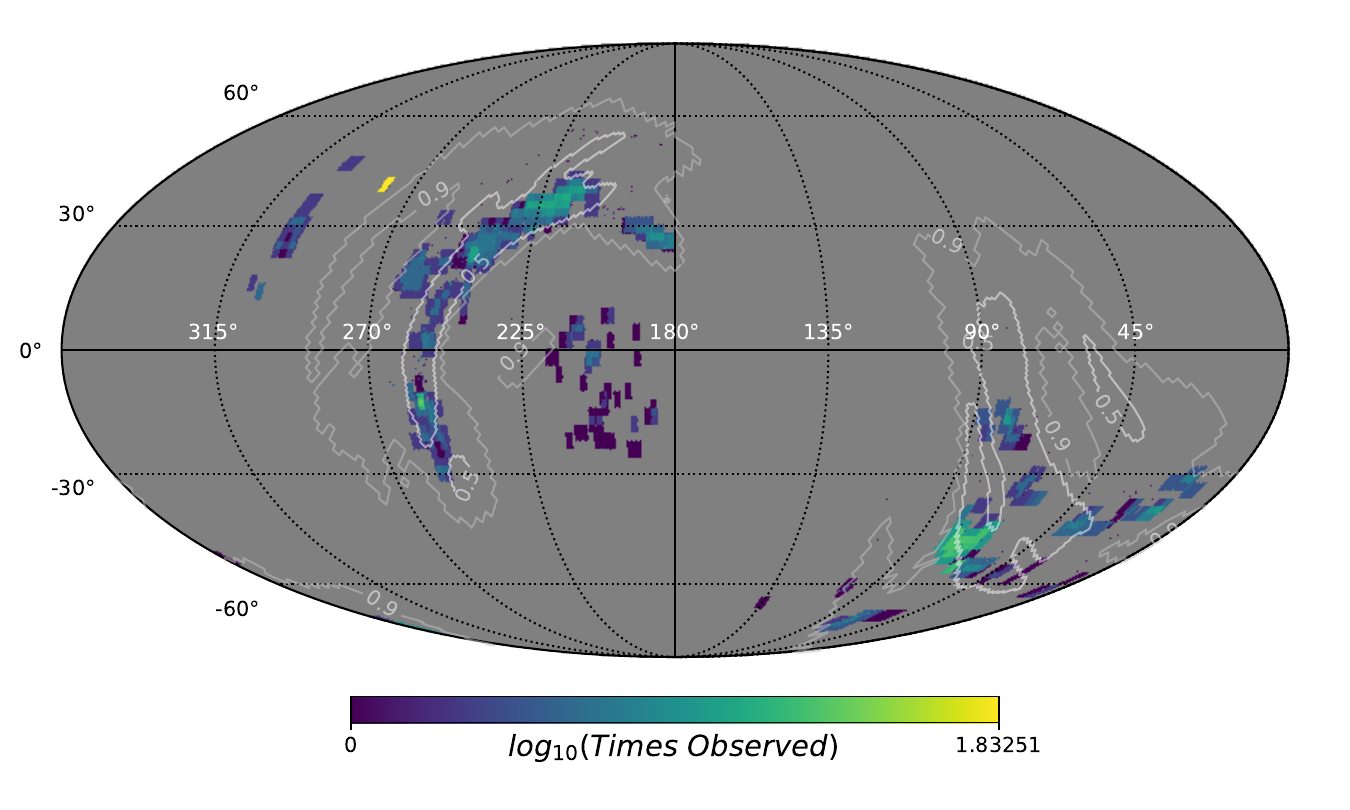}
    \caption{Same as Figure \ref{fig:observations_map}, but separated to northern (top) and southern (bottom) facilities (SWIFT/UVOT coverage is included in both panels).}
    \label{fig:observations_map_split}
\end{figure}

From Figure \ref{fig:observations_map}, it is clear that most of the coverage was obtained in the northern hemisphere, even though the visible area and probability do not show any significant preference to either hemisphere (Table \ref{tab:visibility_by_hemishpere}). We wish to check if this could be due to the different amount of northern observatories and coverage compared to southern ones. Of the ground facilities in Table \ref{tab:observations_summary}, nine are in the southern hemisphere (1M2H/Swope, GRANDMA/LesMakes T60, 1M2H/ANDICAM-CCD, 1M2H/ANDICAM-IR, TAROT TRE, DECam, MPG/GROND, ANU/SkyMapper, and VISTA/VIRCAM) and two (MASTER-Net and LCO 1m) have facilities in both hemispheres. The coverage achieved by the northern (southern) facilities is shown at the top (bottom) panel of Figure \ref{fig:observations_map_split}. We treat the space-based Swift/UVOT facility as a dual-hemisphere facility. 

\begin{figure}[h] 
    \centering
    \includegraphics[width=0.45\textwidth]{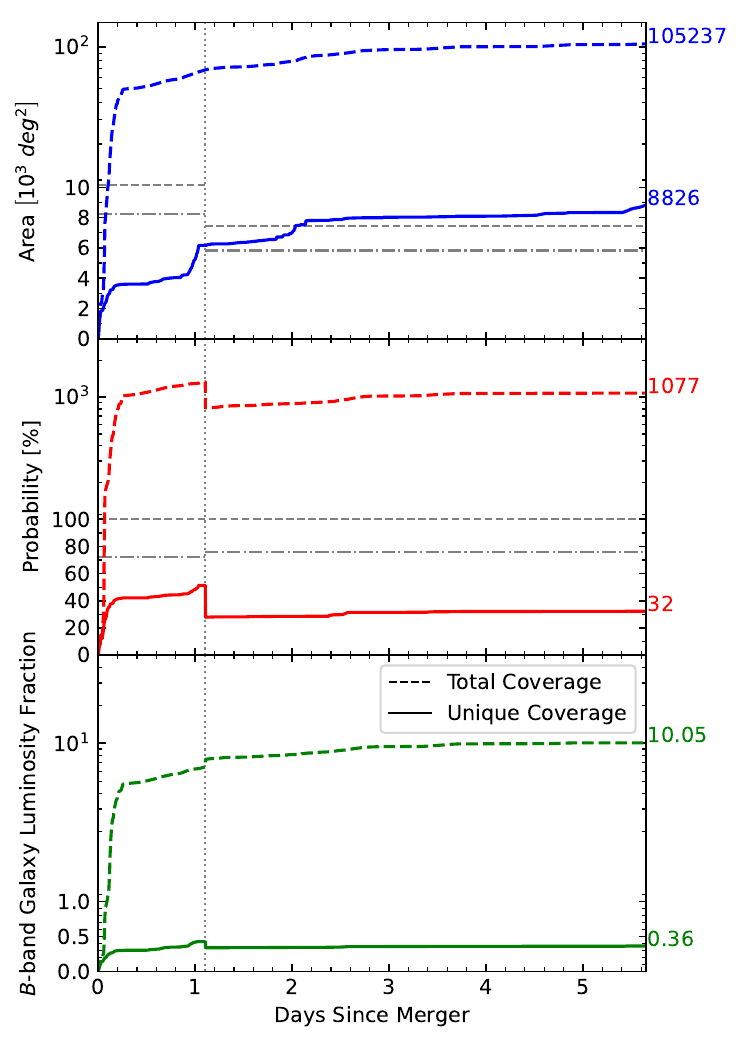}
    \includegraphics[width=0.45\textwidth]{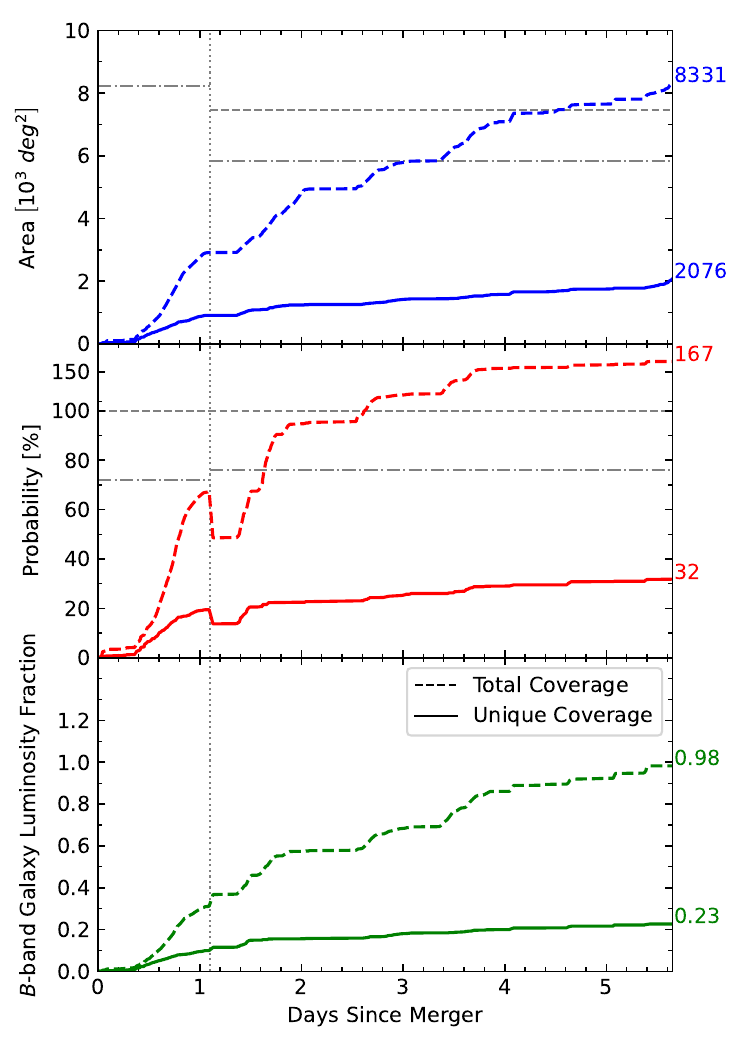}
    \caption{Same as Figure \ref{fig:area_prob_lum_calc}, but separated to northern (left) and southern (right) facilities (SWIFT/UVOT coverage is included in both panels).}
    \label{fig:area_prob_lum_split}
\end{figure}

We repeat our analysis of Sections \ref{subsec:area_and_probability_coverage} and \ref{subsec:galaxy_luminosity_coverage} and examine how the northern (southern) part was covered by northern (southern) facilities. As shown in Figures \ref{fig:observations_map_split} and \ref{fig:area_prob_lum_split}, the majority of the coverage was indeed obtained by northern facilities. However, the southern facilities still contributed to the overlapping coverage in the north, instead of reaching full coverage. Had the southern facilities focused their effort on the southern localization region, while not repeating observations done in the northern localization region, the observable southern part could have been fully covered within 1--2 days.

\section{Additional Kilonova Model Predictions for GW190425}
\label{appendix:additional_models}
We present here additional kilonova models to those shown in Section \ref{subsec:limiting_magnitudes}: the \cite{bulla2019_possis} and \cite{wollaeger2021_supernu} models, and a ``worse case scenario'' for the \cite{nicholl_models}  model.

\begin{figure}[h] 
\includegraphics[width=\textwidth]{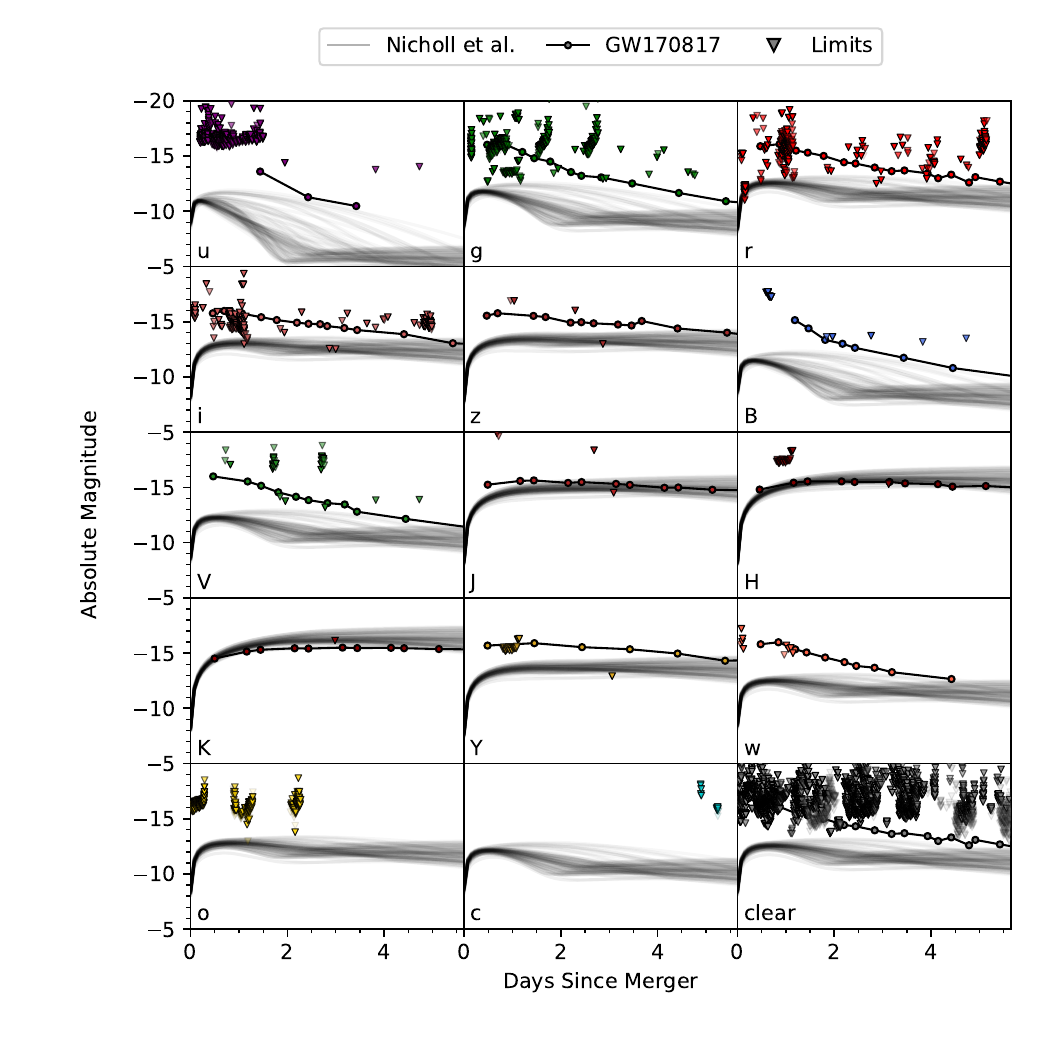}
\caption{Same as Figure \ref{fig:magnitude plots}, but with a ``worst-case'' \cite{nicholl_models} model ensemble.}
\label{fig:mag_plots_worse_case}
\end{figure}

For the \cite{nicholl_models} worse case scenario models (Figure \ref{fig:mag_plots_worse_case}), we use a mass ratio of $0.4<q<0.8$ and a cocoon opening angle of $\cos \theta_{cocoon}=1.0$. The rest of the parameters are the same as those in Table \ref{tab:mosfit_priors}. This results in a redder and dimmer kilonova prediction with the following ejecta parameters for the red ejecta: $M_{ej,red}=0.0543\pm0.0043M_{\odot}$ and $v_{k,red}=0.2628\pm0.0020c$, and for the purple ejecta: $M_{ej,purple}=0.00275\pm0.00041M_{\odot}$, $v_{k,purple}=0.0453\pm0.0010c$ and $\kappa_{ej,purple}\simeq5.6\ cm^2/g$, with no blue component. The total ejecta mass of $M_{ej}=0.0570\pm0.0041M_{\odot}$ has an opacity of $\kappa_{ej}\simeq9.5\ cm^2/g$. Such a kilonova would have been marginally detectable with some of the observations obtained.

\begin{figure}[h] 
\includegraphics[width=\textwidth]{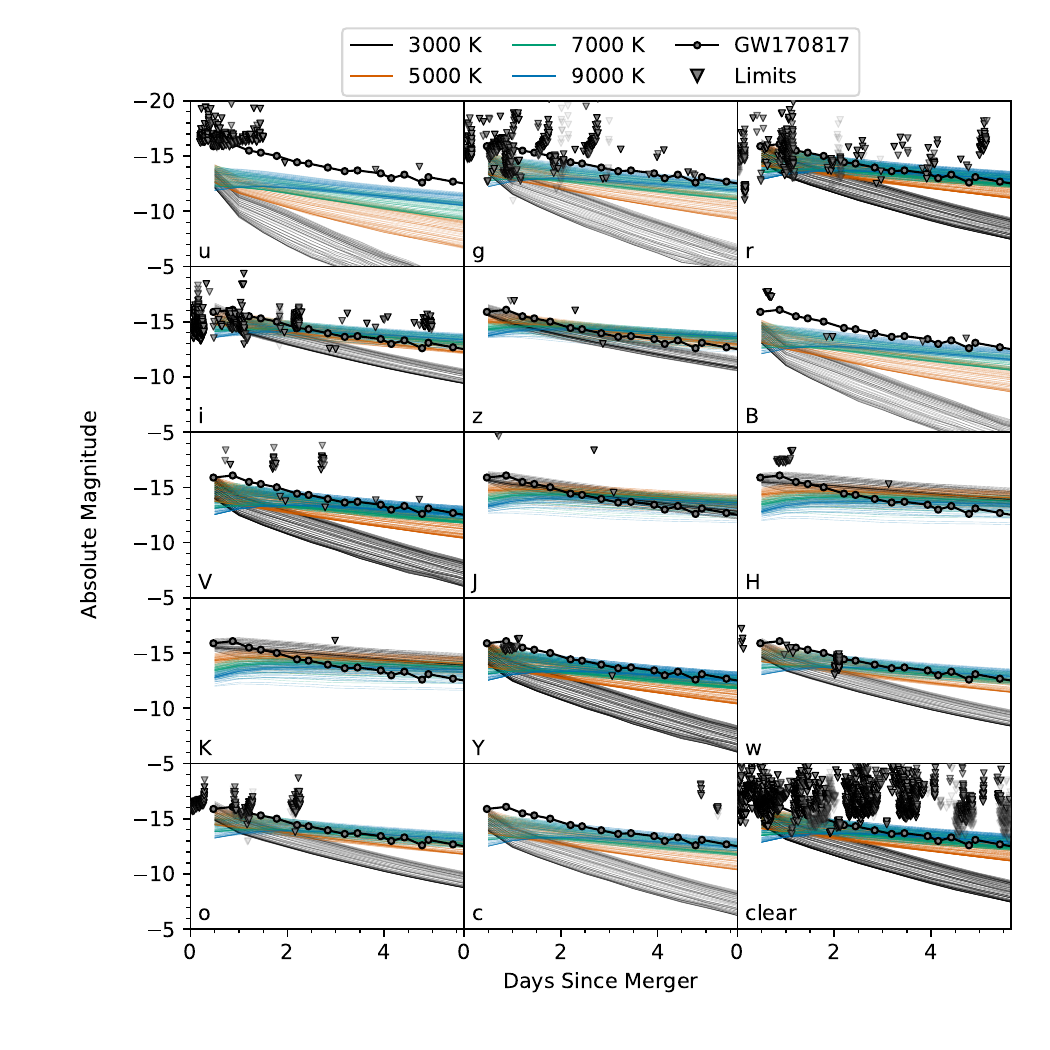}
\caption{Same as Figure \ref{fig:magnitude plots}, but with a suite of {\tt{POSSIS}} models by \cite{bulla2019_possis}. The black, orange, green, and blue lines are for models with initial temperatures of 3000 K, 5000 K, 7000 K, and 9000 K, respectively. The variance in each of the model is due to changes in $\Phi$ and $\theta$, see text for detail.}
\label{fig:mag_plots_possis}
\end{figure}

\cite{bulla2019_possis} use the radiative transfer code POlarization Spectral Synthesis In Supernovae \citep[{\tt{POSSIS}};][]{bulla2015_possis}. The model assumes a two component, axially symmetric geometry, with a red (low $Y_{e}$) ejecta opening angle $\Phi$ above and below the system's equator. The viewing angle is denoted by $\theta$. Given an initial total ejecta mass $M_{ej}$, $\Phi$, and initial temperature $T_{0}$, the code calculates the light curve by evolving the velocities, densities, temperatures, and the opacities of the ejecta. The velocities of the low (high) $Y_{e}$ ejecta range from $0.1c$ to $0.6c$ (from $0.02c$ to $0.1c$). The opacities are wavelength dependent and time dependent, and range from $\kappa=1-10^{2}$ cm$^{2}$/g$^{-1}$ for the low $Y_{e}$ component, and $\kappa=10^{-3}-10^{2}$ cm$^{2}$/g$^{-1}$ for the high $Y_{e}$ component \citep{tanaka2019}. Here we used $M_{ej}=0.04\ M_{\odot}$, similar to the ejecta assumptions in Table \ref{tab:model_results}, while varying $\Phi$, $\theta$, and $T_{0}$ with the following values: $\Phi=[15^{\circ}, 30^{\circ}, 45^{\circ}, 60^{\circ}, 75^{\circ}]$, $\theta=[0^{\circ}, 26^{\circ}, 37^{\circ}, 46^{\circ}, 53^{\circ}, 60^{\circ}, 66^{\circ}, 73^{\circ}, 78^{\circ}, 84^{\circ}, 90^{\circ}]$, and $T_{0}=[3000\ K,\ 5000\ K,\ 7000\ K,\ 9000\ K]$. Figure \ref{fig:mag_plots_possis} shows that while $\Phi$ and $\theta$ have some effect of the brightness of the kilonova, the main factor that affects its detectability is $T_{0}$, especially at late times, with a colder (3000 K) transient being almost non-detectable by the given limits.

\begin{figure}[h] 
\includegraphics[width=\textwidth]{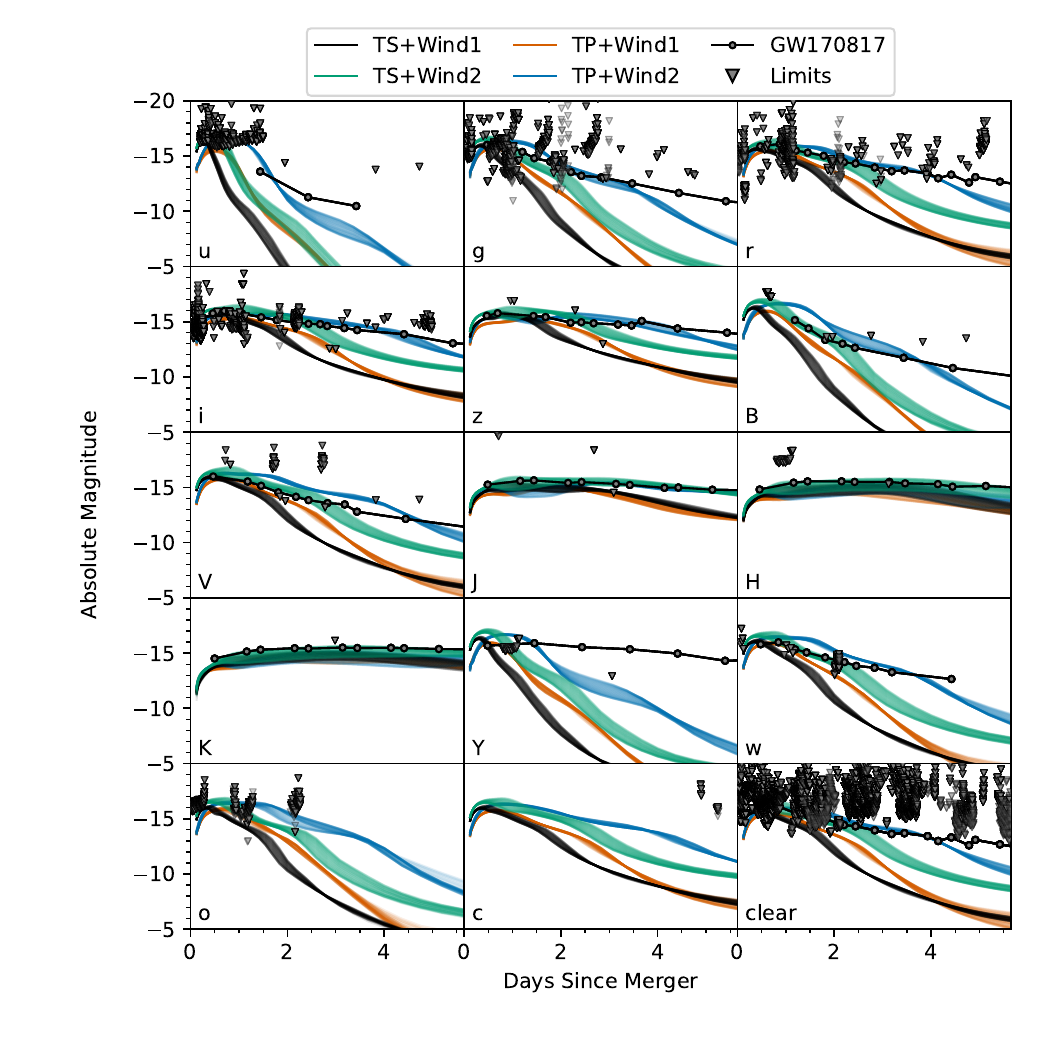}
\caption{Same as Figure \ref{fig:magnitude plots}, but with a suite of {\tt{SuperNu}} models by \cite{wollaeger2021_supernu}. The black, orange, green, and blue lines are for the morphology-content variations of TS+Wind1, TP+Wind1, TS+Wind2, and TP+Wind2, respectively. The variance in each of the model is due to changes in $\theta$. See text for details.}
\label{fig:mag_plots_supernu}
\end{figure}

\cite{wollaeger2021_supernu} use the radiative transfer code {\tt{SuperNu}} \citep{wollaeger2013_supernu, wollaeger2014_supernu} to calculate the light curve from a two component ejecta with axially symmetric geometry, with varying assumptions on the morphology and $Y_{e}$ content of the high $Y_{e}$ component. While the low $Y_{e}$ component morphology is toroidal with $Y_{e}=0.05$, the morphology of the high $Y_{e}$ component can be either spherically symmetric (TS) or have a two polar lobes "peanut" shape (TP). The $Y_{e}$ can either be $Y_{e}=0.37$ (wind1) or $Y_{e}=0.27$ (wind2). To remain consistent with the \cite{foley_updated_parameters} assumptions of a prompt collapse scenario, we use an ejecta mass of $M_{ej}=0.01\ M_{\odot}$ ($0.03\ M_{\odot}$) and velocity of $v_{ej}=0.15c$ ($0.15c$) for the low (high) $Y_{e}$ component, while varying the morphology and high $Y_{e}$ content. The opacities are calculated based on the line-binned approach by \cite{fontes2020}. These models take into account the viewer's observing angle $\theta$, on 54 angular bins. Figure \ref{fig:mag_plots_supernu} shows that the detectability of the kilonova is highly dependent on the morphology of the ejecta, with the low $Y_{e}$ component more likely to obscure the high $Y_{e}$ component for the TS cases.

\section{Analysis without ZTF and Pan-STARRS observations}
\label{appendix:no_ztf_and_ps_analysis}
\begin{figure}[h] 
    \centering
    \includegraphics[width=0.8\textwidth]{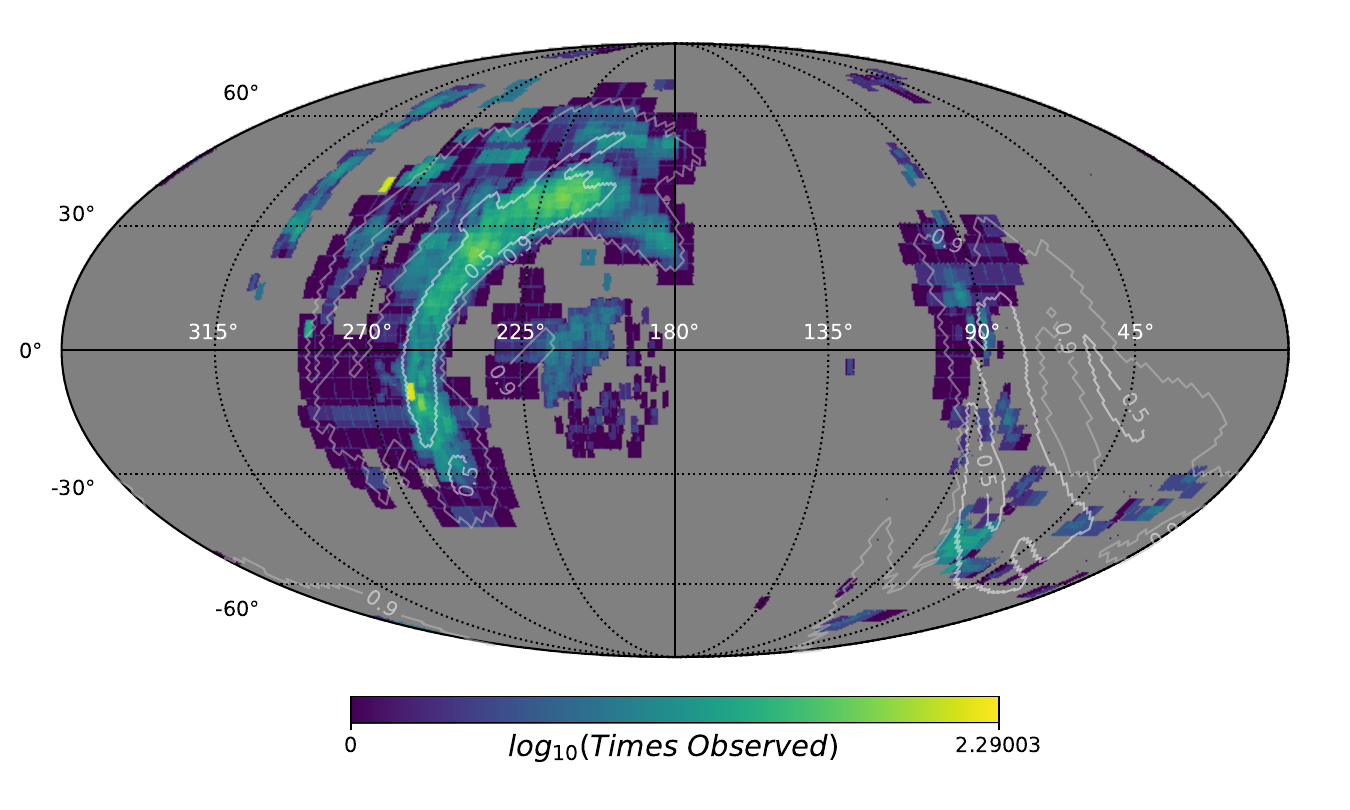}
    \includegraphics[width=0.8\textwidth]{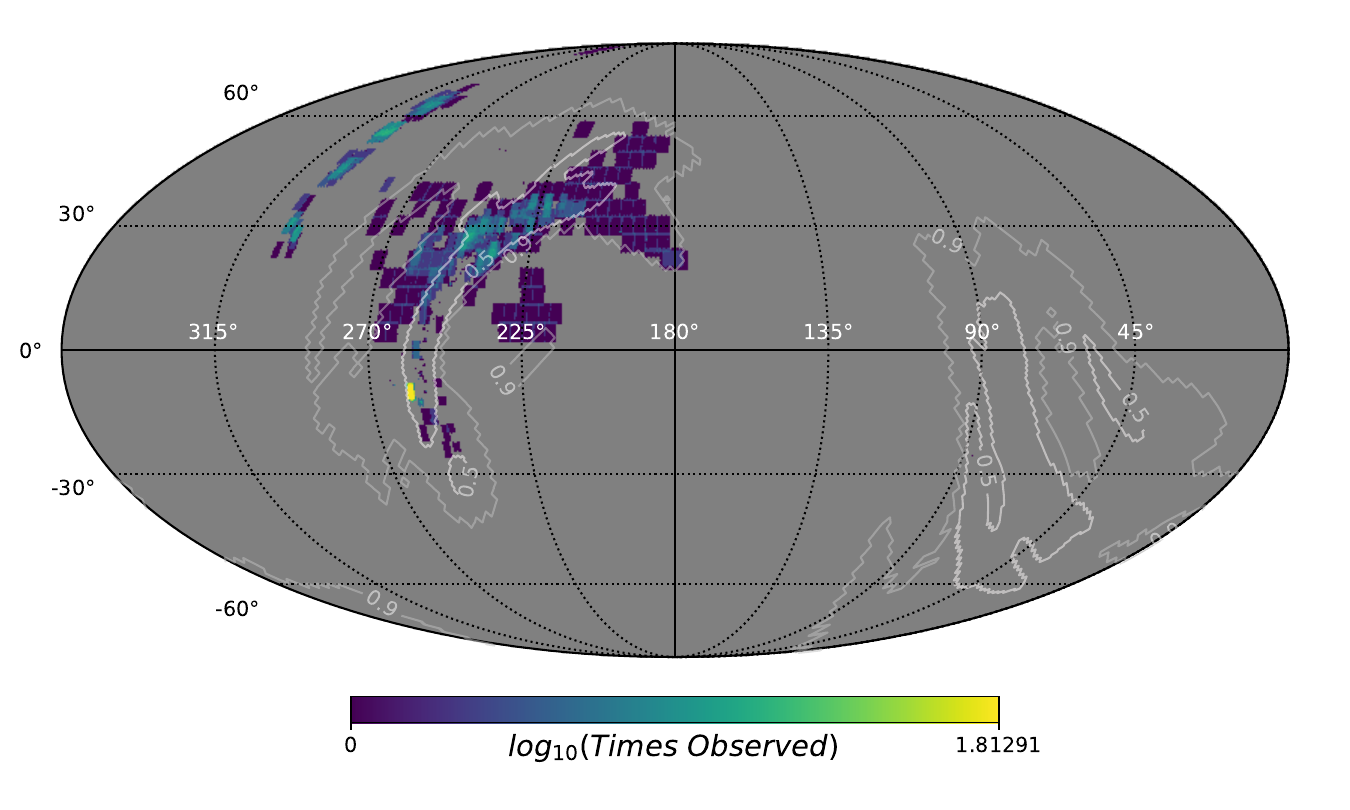}
    \caption{Same as Figures \ref{fig:observations_map} (top; all observations) and \ref{fig:coverage_map_under_gw170817} (bottom; only observations deep enough to detect a GW170817-like kilonova) without the contributions of ZTF and Pan-STARRS.}
    \label{fig:observations_map_no_ztf_no_ps}
\end{figure}

\begin{figure}[h] 
    \centering
    \includegraphics[width=0.45\textwidth]{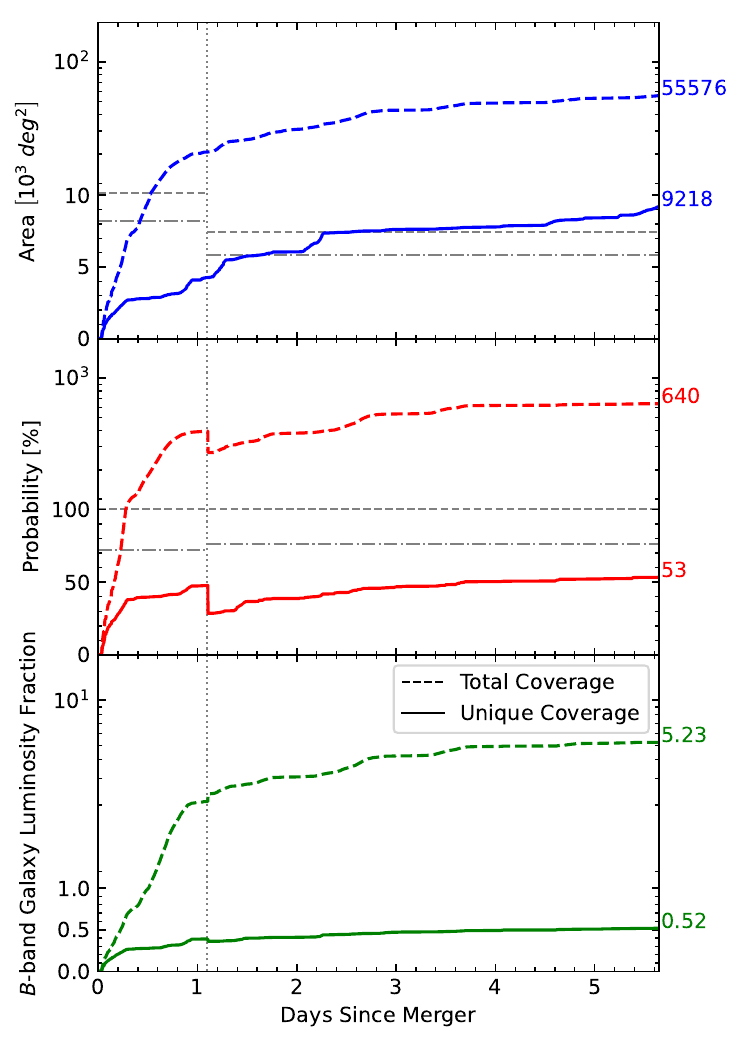}
    \includegraphics[width=0.45\textwidth]{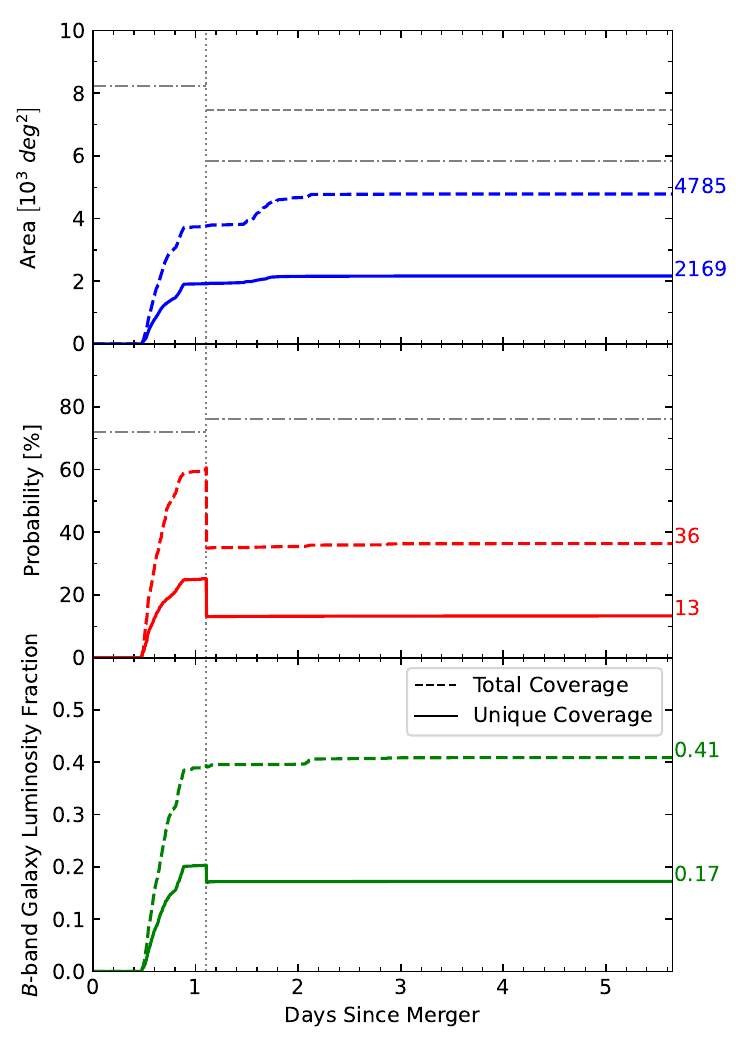}
    \caption{Same as Figures \ref{fig:area_prob_lum_calc} (left; all observations) and \ref{fig:coverage_under_gw170817} (right; only observations deep enough to detect a GW170817-like kilonova) without the contributions of ZTF and Pan-STARRS.}
    \label{fig:area_prob_lum_calc_no_ztf_no_ps}
\end{figure}

We wish to check whether the results of this analysis are driven by one or two dominant surveys in terms of their field of view and depth. We thus repeat our analysis without the ZTF and Pan-STARRS observations. The coverage maps with no ZTF and Pan-STARRS fields are shown in Figure \ref{fig:observations_map_no_ztf_no_ps}. The total coverage in the left panel of Figure \ref{fig:area_prob_lum_calc_no_ztf_no_ps} shows that the full area, probability and galaxy luminosity could have been covered in principle in less than a day even without ZTF and Pan-STARRS. Taking into account only observations that were deep enough to detect a GW170817-like kilonova, more than a third of the localization area, probability and galaxy luminosity could have been covered, as seen in the right panel of Figure \ref{fig:area_prob_lum_calc_no_ztf_no_ps}. Even in this case, coordination could have increased the area, probability, and galaxy luminosity covered by more than a factor of 2 compared the uncoordinated observations.

\begin{figure}[h] 
    \centering
    \includegraphics[width=\linewidth]{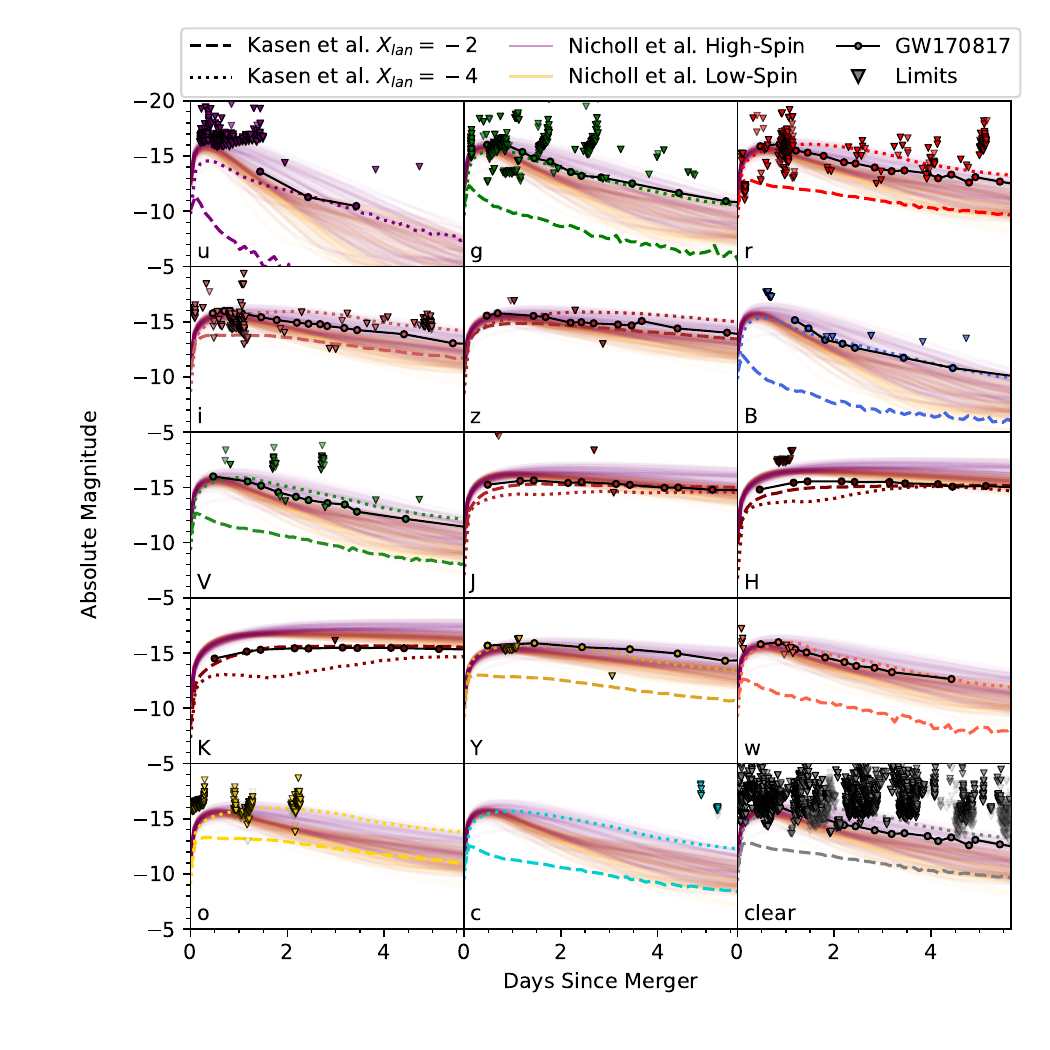}
    \caption{Same as Figure \ref{fig:magnitude plots} without the contributions of ZTF and Pan-STARRS.}
    \label{fig:no_ztf_and_ps_magnitudes}
\end{figure}

In Figure \ref{fig:no_ztf_and_ps_magnitudes} we repeat Figure \ref{fig:magnitude plots} without ZTF and Pan-STARRS, and show that the remaining facilities could have still seen many of the predicted kilonovae associated with GW190425. 
We conclude that our results are not entirely driven by ZTF and Pan-STARRS, and that coordination is important even in the presence of a few dominating surveys.

\end{appendix}

\end{document}